\documentclass[prapplied,twocolumn, superscriptaddress]{revtex4-2}
\usepackage{graphicx}
\usepackage{amssymb}
\usepackage{amsfonts}
\usepackage{amsmath}
\usepackage{lmodern}
\usepackage{mathrsfs}
\usepackage{mathdots}
\usepackage{collref}
\usepackage[colorlinks=true, citecolor=blue, urlcolor=blue]{hyperref}
\usepackage{tikz}
\usepackage{bm}
\usepackage{lipsum} 

\begin{document}

\title{Enhancing low-temperature quantum thermometry via sequential measurements}

\author{Ning Zhang}
\affiliation{School of Materials and Energy, Electron Microscopy Centre of Lanzhou University, Lanzhou University, Lanzhou 730000, China}
\affiliation{Lanzhou Center for Theoretical Physics, Key Laboratory of Quantum Theory and Applications of MoE, and Key Laboratory of Theoretical Physics of Gansu Province, Lanzhou University, Lanzhou, Gansu 730000, China}

\author{Chong Chen}
\email{chongchenn@gmail.com}
\affiliation{Department of Physics, The Chinese University of Hong Kong, Shatin, New Territories, Hong Kong, China}

\author{Ping Wang}
\email{wpking@bnu.edu.cn}
\affiliation{College of Education for the Future, Beijing Normal University, Zhuhai 519087, China}

\begin{abstract}
We propose a sequential measurement protocol for accurate low-temperature estimation. The resulting correlated outputs significantly enhance the low temperature precision compared to that of the independent measurement scheme. This enhancement manifests a Heisenberg scaling of the signal-to-noise ratio for small measurement numbers $N$. Detailed analysis reveals that the final precision is determined by the pair correlation of the sequential outputs, which produces a dependence $N^2$ on the signal-to-noise ratio.  Remarkably, we find that quantum thermometry within the sequential protocol functions as a high-resolution quantum spectroscopy of the thermal noise, underscoring the pivotal role of the sequential measurements in enhancing the spectral resolution and the temperature estimation precision. Our methodology incorporates sequential measurement into low-temperature quantum thermometry, representing an important advancement in low-temperature measurement.
\end{abstract}
\maketitle

\section{Introduction} Accurate temperature measurement of ultracold systems is crucial for quantifying and controlling ultracold atoms, which are central to the study of quantum many-body physics and the development of advanced quantum techniques \cite{Gross2017, Browaeys2020, Ebadi2021, Morgado2021}. The standard measurement method is based on the time-of-flight technique, which, however, is destructive \cite{Leanhardt2003, Gati2006}. Many nondestructive measurement techniques have been proposed \cite{Correa2017, Mehboudi2019, Mitchison2020, Bouton2020, Adam2022}. An extreme example is to map the temperature estimation to a phase estimation problem, where the temperature is measured using a Ramsey interferometry \cite{Stace2010, Johnson2016, Razavian2019, Yuan2023}. Without heat exchange between the thermometer and the sample, the phase estimation scheme provides an almost nondestructive way to detect temperature \cite{Mitchison2020, Adam2022}.

In the low-temperature regime, the quantum thermometer encounters the error divergence problem \cite{Mehboudi2019b, Potts2019, Jorgensen2020}, which arises due to the vanishing heat capacity of the thermometer in the ultracold scenario. Various quantum features have been proposed to improve the precision of measurement, such as strong coupling \cite{Correa2017, Mehboudi2019, Mihailescu2023, Brenes2023}, quantum correlations \cite{Seah2019, Alves2022}, quantum criticality \cite{Hovhannisyan2018, Mirkhalaf2021, Aybar2022, Zhang2022}, and quantum non-Markovianity \cite{Zhang2021, Xu2023}.  Among them, the correlation between the thermometer and the sample is a crucial resource for low-temperature quantum thermometry \cite{Zhang2023}. The sample-induced correlation between different thermometers in the parallel strategy has been shown to be helpful in improving the precision of low temperature \cite{Planella2022,Brattegard2024, Zhang2024}. Beyond the parallel strategy, sample-induced correlations are also present within a sequential protocol between sequential outputs \cite{Burgarth2015, DePasquale2017, Seah2019, Montenegro2022,Radaelli2023}. The quantum thermometry based on the sequential protocol has been proposed in Refs. \cite{DePasquale2017,Seah2019}, however, a negative conclusion is obtained that there is no precision enhancement compared to the independent protocol \cite{DePasquale2017}.

In stark contrast to parallel strategies that take advantage of spatial correlation or entanglement \cite{Giovannetti2006, Giovannetti2011}, the sequential protocol utilizes temporal correlation or coherence to improve measurement precision \cite{Burgarth2015, Braun2018}.  Sequential protocols have been used to measure quantum nonlinear spectroscopy \cite{Wang2019, Wang2021,  Wu2024, Cheung2024, Meinel2022,Shen2023}, which exceeds the resolution limitations set by the lifetime of the probe, allowing for a high spectral resolution \cite{Laraoui2013, Aslam2017, Boss2017, Glenn2018, Pfender2019, Cujia2019}.  Nevertheless, the potential benefits conferred by temporal correlation within the domain of quantum thermometry have yet to be observed.

In this paper, we study the low-temperature quantum thermometry in a sequential protocol. Our goal is to investigate the role of temporal correlation in low-temperature quantum thermometry.  We find that precision can be significantly enhanced by sequential measurements in the regime $\hbar \beta/t_2 \gg 1$, where $\beta=1/K_B T$ denotes the thermodynamic beta and $t_2$ denotes the coherence time of the thermometer.   A typical example is a pK sample measured with a microsecond coherence time thermometer, which yields $\hbar \beta /t_2\approx 7\times 10^3$.   Specifically, a Heisenberg scaling is achieved when the number of sequential measurements satisfies $N  \ll \hbar \beta/t_2$. On the other hand, when $N \gg \hbar \beta/t_2$, this scaling changes to the standard quantum limit, but with a significant improvement in precision compared to the independent measurement scheme. Detail analysis shows that pair correlations in these measurement outputs play a central role in precision enhancement. Given that the pair correlation is intrinsically linked to the noise spectrum of the thermal sample, we elucidate that the temperature estimation process via sequential measurement is fundamentally intertwined with the high-resolution spectroscopy of the noise.

\begin{figure}[htp]
    \centering
    \includegraphics[width=0.95\columnwidth]{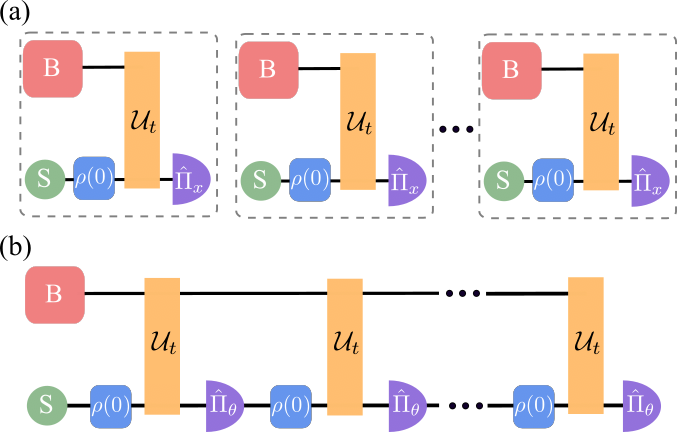}
    \caption{Independent measurement scheme (a) and sequential measurement scheme (b) for temperature estimation. The thermal sample and the thermometer are denoted as ``B" and ``S", respectively. Each block in (a) indicates a temperature estimation process. Here, $\rho(0)$ denotes the initial state preparation, $\mathcal{U}_{t}$ denotes the evolution of the thermometer-sample system, and $\hat{\Pi}_{x}$ and $\hat{\Pi}_{\theta}$ indicate the measurement along $x$ and $\vec{e}_{\theta}:=\cos\theta \vec{e}_{x}+\sin\theta \vec{y}$ direction, respectively.}
    \label{fig:Scheme}
\end{figure}

\section{low-temperature quantum thermometry} We consider a temperature estimation process based on Ramsey interferometry. The thermometer is modeled as a two-level system, and the thermal sample is represented as a multimode bosonic reservoir with a specific temperature $T$. The Hamiltonian of the thermometer in the thermal sample reads $\hat{H}(\tau)= \hat{X}(\tau)\hat{\sigma}_z$  \cite{Mitchison2020,Adam2022},  where $\hat{\sigma}_z$ is the Pauli matrix and $\hat{X}(\tau)$ denotes the noise operator of the sample with the following expression
\begin{equation}    \label{eq:noiseFiled}
    \hat{X}(\tau)=\sum_k \hbar g_k(\hat{b}^{\dagger}_k e^{i\omega_k \tau}+h.c.)+\hat{\zeta}(\tau).
\end{equation}
Here $\hat{b}^{\dagger}_k$ $(\hat{b}_k)$ denotes the creation (annihilation) operator of the $k$-th noise mode with frequency $\omega_k$ and coupling strength $g_k$ and $\hat{\zeta}(\tau)$ characterizes the other possible noises caused by temperature-independent mechanisms.  Furthermore, $\hat{\zeta}(\tau)$ is assumed to be Gaussian with zero mean and the noise operator $\hat{b}_k$ satisfies $\langle \hat{b}_k \rangle_{B}=0$ and $\langle \hat{b}^{\dagger}_k \hat{b}_k^\prime \rangle_{B} = \bar{n}_k \delta_{k,k^\prime}$, where $\langle \rangle_B$ denotes the average over the state of the sample and $\bar{n}_k=(e^{\beta \hbar \omega_k}-1)^{-1}$. In this study, we focus on the low temperature regime with $\hbar \beta/t_2 \gg 1$, where $t_2$ denotes the coherence time of the thermometer under the influence of $\hat{X}(\tau)$. This regime can involve a sub-nK or pK sample couples with a $\mu$s or microsecond thermometer \cite{Kovachy2015,Mehboudi2019}. 

Before delving into the discussion of sequential measurements, let us first illustrate the principle of temperature estimation based on independent measurement \cite{Razavian2019,Zhang2023}, as shown in Fig. \ref{fig:Scheme} (a). The thermometer is initially prepared in the $x$ direction, denoted $\rho(0)=|+\rangle \langle +|$. Then it interacts with the sample and undergoes a free evolution, denoted as $\mathcal{U}_t$. After time $t$, a projective measurement $\hat{\Pi}_{x;s}$ along the $x$ axis is applied to the thermometer with the output $s=\pm 1$ and the corresponding probability $P_{s}$.   Repeat this measurement process $N$ times, as shown in Fig. \ref{fig:Scheme} (a).  The temperature is estimated from these outputs. The final precision of the temperature is bounded by the Cram\'{e}r-Rao bound
\begin{equation}\label{eq:CRB}
  (\beta/\Delta \beta)^2_{ind} \le N \beta^2 \mathcal{F} \text{ with } \mathcal{F}=\sum_{s=\pm} P_{s} \mathcal{L}_{\beta}^2, 
\end{equation}
where  $\left(\beta/\Delta \beta \right)^2_{ind}$ denotes as the quantum signal-to-noise ratio (QSNR), $\mathcal{F}$ denotes the Fisher information, and $\mathcal{L}_{\beta}=-\frac{d \ln P_{s}}{d \beta}$ is known as the score function.

For the dephasing process, the probabilities $P_{\pm}$ can be solved exactly \cite{Breuer2007} with the results $P_{\pm}=(1 \pm e^{-\Gamma(t)})/2$,  where $\Gamma(t)= \frac{2}{\hbar^2}\int^{t}_{0}d\tau_1 \int^{t}_0 d\tau_2 \langle \{\hat{X}(\tau_1), \hat{X}(\tau_2)\} \rangle_B $ with $ \{\hat{A}, \hat{B}\} \equiv \frac{1}{2}(\hat{A}\hat{B}+\hat{B}\hat{A})$.  The QSNR is then obtained as
\[ (\beta/\Delta \beta)^2_{ind} \le \frac{4\beta^2 D^2_{0}}{e^{2 \Gamma (t)}-1}N,\]
where $D_0 \equiv  \frac{1}{2}\frac{d}{d-\beta} \Gamma(t)=4\hbar \int d\omega \frac{J(\omega)}{\omega} \bar{n}(\omega)(1+\bar{n}(\omega))(1-\cos \omega t)$ with $J(\omega)\equiv \sum_k g^2_k \delta(\omega-\omega_k)$ being the spectral density \cite{Razavian2019}.  In the low-temperature regime,  only low-frequency modes in the sample have nonzero contributions to temperature estimation as $\bar{n}(\omega)\approx 0$ when $\beta \hbar \omega \gg 1$.  Setting the evolution time as the coherence time $t_2$, determined from the condition $\Gamma(t_2)=1$ and employing the approximation $1-\cos\omega t_2\approx \frac{1}{2} \omega^2 t^2_2$  for low-frequency modes,  we get 
\begin{equation}\label{eq:precision-IM}
 (\beta/\Delta \beta)^2_{ind} \le \frac{g^4 t^4_2}{e^2-1}N,
\end{equation}
where $g^2  = 4\hbar \beta\int^{\infty}_0 d\omega \omega J(\omega) \bar{n}(\omega)[1+\bar{n}(\omega)]$ characterizes the effective coupling strength between the  low-frequency noise and the thermometer.   It shows that both $g$ and $t_{2}$ serve as crucial resources to improve the accuracy of the estimation.  Laudau bound $\left(\beta/\Delta \beta\right)^2_{ind} \approx N$ is achieved when $g t_2 \approx 1$.  However, this condition cannot be satisfied in ultra-cold systems, since $t_2 \sim const $ while $g\sim 0 $ \cite{Jorgensen2020}.  For example, when $\hbar \beta /t_2 \gg1$, we have $g^2 t_2^2 \approx (\hbar \beta/t_2)^{-(s+1)}\ll 1$ for the Ohmic class spectral $J(\omega)=\alpha \omega^{s} \omega^{1-s}_c e^{-\omega/\omega_c}$ with $s> 0$. 

Unlike the dependence $t^2_2$ observed in magnetic field sensing \cite{Degen2017}, the temperature estimation is outlined in Eq. \eqref{eq:precision-IM} reveals a  $t^4_2$ dependence. The discrepancy arises from the intrinsic nature of the temperature, which is not directly related to the single integral of $\langle \hat{X}(\tau) \rangle_B$ but to the double integrals of the two-point correlation function $\langle \{\hat{X}(\tau), \hat{X}(\tau') \}\rangle_B$ \cite{Zhang2024}. Generally, one can decompose $t^4_{2}$ as $t^2_2 \times t^2_2$, where the first $t^2_2$ term denotes the coherent time of the thermometer, as observed in the case of magnetic field sensing, while the other $t^2_2$ term represents the correlation time $\max |\tau-\tau'|$ in the two-point correlation function. For independent measurements, both the coherence time and the correlation time are limited to $t_2$.  Extending both timescales can improve the accuracy of the temperature estimation. Given a thermometer, the extension of the coherent time presents significant challenges, while increasing the correlation time is relatively easy for low-temperature estimation.   In the following, we will illustrate how the correlation time is enlarged by sequential measurements.

\section{Sequential measurement scheme}
 The sequential measurement scheme is shown in Fig. \ref{fig:Scheme} (b), where the independent measurement is replaced by a series of continuous measurements. In each measurement process, the thermometer is still prepared in the $|+\rangle \langle +|$ state, while the measurement axis is along the $\vec{e}_{\theta}=\cos \theta \vec{x} + \sin \theta \vec{y}$ direction.  Taking into account each initial state preparation and the following projective measurement, the readout probability $P_{s_N, \cdots,s_1}$ for $N$ sequential measurements is derived as
\begin{equation}\label{eq:prob-Corr}
    P_{\bm{S}}=\langle \mathcal{M}_{s_N} \cdots \mathcal{M}_{s_2} \mathcal{M}_{s_1}\rangle_B,
\end{equation}
where $\bm{S}\equiv s_N,\cdots,s_1$ and $\mathcal{M}_{s_j}\equiv\text{Tr}_S[\hat{\Pi}_{\theta;s_j}\mathcal{U}_{t}\rho(0))]$ represents the positive operator-valued measure (POVM) induced by $j$-th Ramsey interference process applied on the sample.  In terms of the $z$-direction basis $\{|0\rangle ,|1\rangle\}$,  the measurement operator $\hat{\Pi}_{\theta;s_j}$ has the following expression
\begin{equation}
    \hat{\Pi}_{\theta;s_j}=\frac{1}{2}\sum^{0,1}_{\eta,\bar{\eta}}e^{i \eta^{-} (\theta+\frac{1-s_j}{2}\pi)}|\eta\rangle \langle \bar{\eta}|
\end{equation}
where $\eta^{\pm}= (\eta-\frac{1}{2})\pm(\bar{\eta}-\frac{1}{2})$.  The detail calculation yields [see Appendix \ref{sec:ep-app} for derivations]
\begin{equation}\label{eq:prob-eta}
    P_{\bm{S}}=\sum_{\{\eta_j,\bar{\eta}_j\}} \mathcal{G}(\bm{\eta},\bar{\bm{\eta}}) \prod_{j}\frac{e^{i \eta^{-}_j (\theta+\frac{1-s_j}{2}\pi)}}{4}, 
\end{equation}
where $\bm{\eta}=\{\eta_{j}\}$ and $\bar{\bm{\eta}}=\{\bar{\eta}_{j}\}$ and
    \begin{align*}\label{eq:propagator-exac}
    \mathcal{G}= & e^{-2\int^{Nt}_{0}d\tau_1 \int^{Nt}_{0}d\tau_2  \eta^{-}(\tau_1)  [C^{++}_{\tau_1,\tau_2}\eta^{-}(\tau_2)+i C^{+-}_{\tau_1,\tau_2}   \eta^{+}(\tau_2)]}
\end{align*}
characterizes the evolution ``probability"  along the specified paths $\bm{\eta}$ and $\bar{\bm{\eta}}$.  Here,
 $C^{++}_{\tau_1,\tau_2}\equiv \frac{1}{\hbar^2}  \langle \{\hat{X}(\tau_1), \hat{X}(\tau_2)\} \rangle_{B}$  and $C^{+-}_{\tau_1,\tau_2}\equiv  -\frac{i}{\hbar^2} \Theta(\tau_1-\tau_2) \langle [\hat{X}(\tau_1), \hat{X}(\tau_2)] \rangle_{B}$  are defined as the classical and quantum correlation of noise, respectively \cite{Wang2019,Wang2021,Cheung2024,Wu2024} with $\Theta(\tau_1-\tau_2)$ a step function, $[\hat{A},\hat{B}]\equiv \hat{A}\hat{B}-\hat{B}\hat{A}$, and $\eta^{\pm}(\tau)\equiv \eta^{\pm}_j$ for $j=\tau/t$.

In the low-temperature scenario, we notice that only low-frequency modes are relevant for temperature estimation. It is appropriate to divide $\hat{X}(\tau)$ into the low-frequency part and the high-frequency part, denoted $\hat{X}(\tau)= \hat{X}_{L}(\tau)+\hat{X}_{H}(\tau)$. Compared with the low-frequency part, the high-frequency part exhibits a relatively short correlation time and dominates the decoherence dynamics of the thermometer. Thus, one can approximate high-frequency noise as a white noise that satisfies $\langle \{\hat{X}_H(\tau) \hat{X}_{H}(\tau^{\prime})\}\rangle=\frac{1}{t_2} \delta(\tau-\tau^{\prime})$.  Under this approximation,  $\mathcal{G}$ is simplified to 
\begin{equation}\label{eq:propagator-Exp}
    \mathcal{G}=e^{- \sum_{j,l}[2C^{++}_{l,j} \eta^{-}_{l}\Delta \eta^{-}_{j}  +2 i  C^{+-}_{l,j}\eta^{-}_l \eta^{+}_j] -\sum_j  (\eta^{-}_j)^2\frac{ t}{t_2}},
\end{equation}
where $C^{+\pm}_{l,j}=\int^{lt}_{(l-1)t}d\tau_1 \int^{jt}_{(j-1)t}d\tau_2 C^{+\pm}_{\tau_1,\tau_2}$ denotes the correlations induced by the low-frequency noise between $l$-th and $j$-th measurements. Here, the identify $\eta^{+}_j \eta^{-}_j=0$ is used. With the assumption that the low-frequency noise induced correlations are very weak, i.e. $|C^{+\pm}_{j,i}|\ll 1$, the detailed expression of $P_{\bm{S}}$ is obtained up to the first order of $C^{+\pm}_{j,i}$ [see Appendix \ref{sec:pd-app} for details].  Result is
\begin{equation}\label{eq:DF}
          P_{\bm{S}}\approx \prod_{i} P_{s_i} \prod_{i<j} [1-s_i s_{j} \frac{\sin^2 \theta  C^{++}_{j,i}}{P_{s_i} P_{s_j}}e^{- \frac{2t}{t_2}}],
\end{equation}
where $P_{s_i}\approx \frac{1}{2}(1+s_i\cos\theta e^{-\frac{t}{t_2}-C^{++}_{i,i}})$.   It reveals that the outputs $\{s_i\}$ remain approximately independent, except for a weak correlation term $\propto C^{++}_{j,i}$. 

By using the probability distribution $P_{\bm{S}}$, the score function, denoted $\mathcal{L}_\beta\equiv -\frac{d}{d\beta} \ln P_{\bm{S}}$,  is derived as
\begin{equation} \label{eq:Score-C}
    \mathcal{L}_{\beta}\approx \sum_{i}\frac{s_i D_{0}}{P_{s_i}} \cos \theta e^{-\frac{t}{t_2}} -\sum_{i<j} \frac{s_i s_j D_{j-i} }{P_{s_i} P_{s_j}}\sin^2\theta e^{-\frac{2t}{t_2}},
\end{equation}
where $D_{j-i} \equiv -\frac{d C^{++}_{j,i}}{d\beta}$. Using further the mode decomposition of $\hat{X}_L(\tau)$, as shown in the Eq. \eqref{eq:noiseFiled}, the detailed expression of $D_{l}$ is obtained as
\begin{equation}\label{eq:Dji}
    D_{l}\approx 2 \hbar t^2 \int d\omega \omega J(\omega)  \bar{n}(\omega)(1+\bar{n}(\omega))\cos l\omega t.
\end{equation}
Here, $\bar{n}(\omega)=(e^{\beta \hbar \omega}-1)^{-1}$ introduces a natural frequency truncation with $\bar{n}(\omega)\ll 1$ when $\beta \hbar \omega \gg 1$. From Eq. \eqref{eq:Score-C}, one can find that the first term captures contributions from independent measurements, while the second term captures the contributions of the correlation between different outputs.  The pairwise property of the correlation term indicates that there are $N(N-1)/2$ independent terms that can contribute to the Fisher information shown in Eq. \eqref{eq:CRB}, which would significantly improve the accuracy of the estimation.  Interestingly, the contributions of these two terms are controlled by the measurement axis $\theta$. When $\theta=0$, it yields the result of independent measurements.  When $\theta=\pi/2$, it serves as a correlation measurement thermometer.

\begin{figure}[thbp]
    \centering
    \includegraphics[width=0.95\columnwidth]{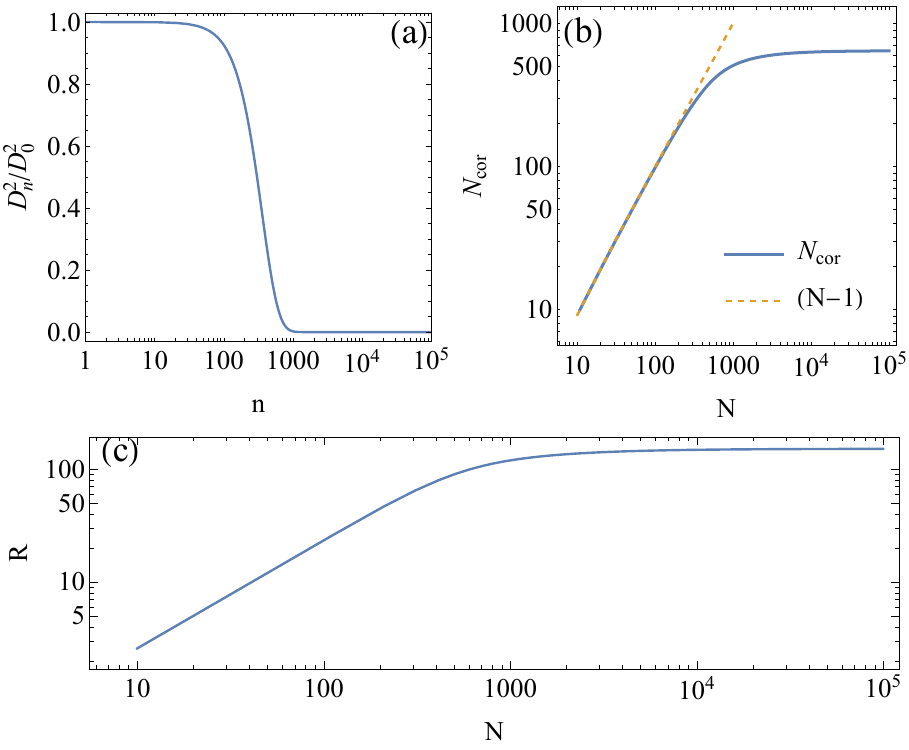}
    \caption{(a) The relative correlation strength $D^2_{n}/D^2_{0}$ decays with distance $n$, yielding the correlation length $N_c\approx 380$.  (b) The change of $N_{cor}$ with the number of measurements $N$.  When $N\ll N_c$, $N_{cor}$ increases linearly (denoted by the orange dashed line). When $N\gg N_c$, $N_{cor}$ saturates to a fixed value $N_s\approx 640$. (c) The enhancement factor $R$ as a function of $N$.  Here, an Ohmic noise spectrum $J(\omega)=\alpha \omega e^{-\omega/\omega_c}$ is taken for low-frequency noise. Parameters are: $\alpha=0.1$, $\omega_c=10.0$,  $\omega_c t_2=1$, and  $\hbar\beta \omega_c /t_2=10^3$.}
    \label{fig:Ncor}
\end{figure}

\section{estimation precision for the sequential measurement} 
Choosing $\theta=\pi/2$, we concentrate on the enhancement of the estimation precision by correlations between different measurements. By setting $t=t_2$, the Fisher information is calculated according to the Eq. \eqref{eq:CRB}.  The QSNR is then determined from the Cram\'{e}r-Rao bound, given by
\begin{equation}\label{eq:precision-CM}
   (\beta/\Delta \beta)^2_{cor}  \le  2 e^{-4} g^4 t^4_2  NN_{cor}, 
\end{equation}
where $N_{cor} \equiv \frac{2}{N}\sum_{i<j}  D^2_{j-i}/D^2_{0}$ and the equality $\beta D_{0}=g^2 t^2_2/2$  is used.  Here, $D^2_{j-i}/D^2_{0}$ quantifies the relative strength of correlation between $i$-th and $j$ -th measurements in the sequence, which is a decay function of $j-i$.  The correlation length $N_c$ can be defined from the condition $D^2_{n}/D^2_{0}=e^{-1}$, beyond which the correlation effect can be safely ignored.  

When $N \ll N_c$, $N_{cor}\approx (N-1)$, which results in a scaling of $N^2$ of the QSNR from Eq. \eqref{eq:precision-CM}
\begin{equation}\label{eq:precision-Heisenberg}
      (\beta/\Delta \beta)^2_{cor}  \le  2 e^{-4} g^4 t^2_4  N (N-1).
\end{equation}
In the low-temperature case, the condition $N_c \sim \hbar \beta/t_2 \gg 1$ suggests that the Heisenberg scaling holds for a wide range of measurement counts $N$. In contrast,  when $N\gg N_c$, $N_{cor}$ would saturate to a fixed value $N_{s} \propto  N_c$, which yields an $N$ scaling of the QSNR, given by
\begin{equation}\label{eq:precision-largeN}
 (\beta/\Delta \beta)^2_{cor}  \le  2 e^{-4} g^4 t^4_2 N  N_s.
\end{equation}
However,  the property $N_s \propto  N_c \sim \hbar \beta /t_2$ indicates that a large enhancement can be expected for low-temperature estimation. We can owe this improvement to the increase in the correlation time in the two-point correlation function $\langle \{\hat{X}(\tau), \hat{X}(\tau')\} \rangle_{B}$, which changes from $t_2$ for independent measurements to $ N_{s}t_2$ for sequential measurements, resulting in $\mathcal{F}_\beta \sim g^4 t^2_2 (N_s t_2)^2 \frac{N}{N_s}$.

To reflect the enhancement of the measurement precision induced by the sequential measurements, we introduce an enhancement factor 
\begin{equation}
    R  \equiv (\Delta \beta)^2_{ind}/(\Delta \beta)^2_{cor}.
\end{equation}
Assuming that the Cram\'{e}r-Rao bound is saturated, one can find that the enhancement factor $R$ increases from $2 (e^{-2}-e^{-4}) (N-1)$ to $2 (e^{-2}-e^{-4}) N_s$ as $N$ increases.

To elucidate the advantage of sequential measurements, we consider the case where low-frequency noise has Ohmic spectral $J(\omega)=\alpha \omega e^{-\omega/\omega_c}$ and temperature-independent noise is ignored.  The results are illustrated in Fig. \ref{fig:Ncor} with the parameters $\hbar \beta/t_2=10^3$. From Fig. \ref{fig:Ncor} (a), one can observe that when $n$ is small, the relative correlation strength $D^2_{n}/D^2_{0}$ is almost constant, while it decays to zero when $n$ is large enough.  Consequently, when $N$ is small, a linear increase of $N_{cor}$ is observed in Fig. \ref{fig:Ncor} (b), which saturates to a fixed value $\sim \hbar \beta/t_2$ when $N$ is large enough.  In correspondence, the enhancement factor shows a linear increase for the case of small $N$, while it saturates to a fixed value $\sim \hbar \beta /t_2$ in the limit of large $N$ .

\section{Quantum spectroscopy}  
One can find that that temperature estimation process is equivalent to measuring the noise spectrum of $\hat{X}(\tau)$. To illustrate this equivalence relation, consider the pair correlation in sequential measurements with the setting $\theta=\pi/2$. The mean value of these pairwise outputs, defined as $S_{j,i}\equiv \langle s_j s_i\rangle-\langle s_j \rangle\langle s_i\rangle$, are obtained by using the explicit expression of $P_{\bm{S}}$ .  The result is
\begin{equation}\label{eq:Correlation-irre}
	S_{j,i}\approx 4 t^2 e^{-2t/t_2}  C^{++}_{j,i}
\end{equation}
One can find that $S_{j,i}$ is directly proportional to the two-point correlation function $C^{++}_{j,i}$. After the Fourier transformation of $S_{j,i}$, the noise spectrum is obtained. Thus, the pair correlation corresponds exactly to the noise spectrum up to a factor $4t^2e^{-2t/t_2}$.  

Note that the spectral resolution obtained from this method is independent of the probe's (thermometer's) coherence time, thus facilitating high-resolution spectroscopy \cite{Boss2017}. These results differ significantly from the noise spectrum obtained by independent measurements, where only $C^{++}_{i,i}$ is accessed, thereby limiting the resolution to the sensor's coherence time $1/t_2$. This advantage makes sequential measurements desirable for high-resolution quantum spectroscopy \cite{Pfender2019}. Here, we use sequential measurements to improve the precision of temperature estimation. Because both the correlation functions $C^{++}_{i,i}$ and $C^{++}_{j,i}$ contain temperature information, sequential measurements produce better performance than independent measurements by taking into account all long-term correlations. Hence, both quantum thermometry and high-resolution quantum spectroscopy use sequential measurements to increase the correlation time of the noise in a sample, thereby improving the estimation precision or spectral resolution. 

\section{conclusions} 
In this paper, we introduce the sequential measurement to low-temperature thermometry. We show that sequential measurement significantly improves measurement precision in the ultracold scenario with $\hbar \beta/t_2 \gg 1$. This enhancement stems from the accounting of long-term correlations of low-frequency noise in the thermal sample. A Heisenberg scaling is verified, which holds for a wide range of measurement numbers $N$ in the ultracold scenario with $ \hbar \beta/t_2 \gg 1$. In contrast, a fundamental limit of the estimation precision is reached when $N$ is large enough, determined by the intrinsic correlation time of the low-frequency noise mode. Additionally, we establish a close link between low-temperature quantum thermometry via sequential measurement and quantum correlation spectroscopy. Both of them use sequential measurements to increase the correlation time of the noise in a sample.  Noting that temperature is a statistical quantity, our work opens the door to precision measurement of statistical quantities through correlation measurements.

\section{Acknowledgment}
N. Z. is supported by the National Natural Science Foundation of China (Grant No. 12247101), the Fundamental Research Funds for Central Universities (Grant No. lzujbky-2024-jdzx06) P. W. is supported by the National Natural Science Foundation of China(Grant No. 1247050444), the Talents Introduction Foundation of Beijing Normal University (Grant No. 310432106), Innovation Program for Quantum Science and Technology of China (Project 2023ZD0300600) and Guangdong Provincial Quantum Science Strategic Initiative (Grant No. GDZX2303005).

\appendix

\section{evolution probability $\mathcal{G}(\eta,\bar{\eta})$} \label{sec:ep-app}

In terms of the basis of the z direction $\{|0\rangle ,|1\rangle \}$, the measurement operator along the $\theta$-direction reads
\[ \hat{\Pi}_{\theta;s}=\frac{1}{2}\sum^{0,1}_{\eta,\bar{\eta}}e^{i \eta^{-} (\theta+\frac{1-s}{2}\pi)}|\eta_{z}\rangle \langle \bar{\eta}_{z}|,\]
where $\eta^{\pm}=(\eta_i+\frac{1}{2})\pm(\bar{\eta}_i+\frac{1}{2})$.
Using this expression, one can get that
\begin{align}\label{eq:POVM}
\mathcal{M}_{s_j} \rho_B=& \sum^{\pm}_{\eta,\bar{\eta}} \frac{1}{4} e^{i \eta^{-}_{j} (\theta+\frac{1-s_j}{2}\pi)}\mathcal{T} e^{-\frac{i}{\hbar} (2\eta_j-1) \int^{jt}_{(j-1)t}d\tau \hat{X}(\tau) } \nonumber \\
&  \rho_B \bar{\mathcal{T}} e^{\frac{i}{\hbar} (2\bar{\eta}_j -1)\int^{jt}_{(j-1)t} d\tau \hat{X}(\tau)}, 
\end{align}
where $\mathcal{T}$ and $\bar{\mathcal{T}}$ denotes the time-ordering and anti-time-ordering.  Based on this result, one can get that 
\begin{align}
    P_{\bm{S}}=&\frac{1}{4^N}\sum_{\bm{\eta},\bar{\bm{\eta}}}  e^{i \sum_j \eta^{-}_{j} (\theta+\frac{1-s_j}{2}\pi)} \mathcal{G}(\bm{\eta},\bar{\bm{\eta}}),\\ 
    \mathcal{G}=&\langle  \bar{\mathcal{T}} e^{\frac{i}{\hbar}  \int^{Nt}_{0} d\tau \hat{X}(\tau) \bar{\eta}'(\tau)}  \mathcal{T}e^{-\frac{i}{\hbar}  \int^{Nt}_{0}d\tau \hat{X}(\tau) \eta'(\tau)} \rangle_{B},
\end{align}
where $\bm{\eta}=\{\eta_i\}$, $\bar{\bm{\eta}}=\{\bar{\eta}_i\}$ and  $\eta'_j(\tau)\equiv 2\eta_j-1$,  $\bar{\eta}'_j(\tau)\equiv 2\bar{\eta}_j-1$ with $i= \tau/t$. Here, $\mathcal{G}(\bm{\eta},\bar{\bm{\eta}})$ denotes the evolution "probability" along the path $\bm{\eta}$ and $\bar{\bm{\eta}}$. 
For Gaussian noise, the average can be exactly solved and the result is
\begin{align}
    \mathcal{G}(\bm{\eta},\bar{\bm{\eta}})=&  e^{-\frac{1}{\hbar^2} \int^{N t}_0 d\tau_1\int^{t}_{\tau_1} d\tau_2 \bar{\eta}'(\tau_1)\bar{\eta}'(\tau_2) \langle \hat{X}(\tau_1)  \hat{X}(\tau_2)\rangle_B} \nonumber \\
  &   e^{-\frac{1}{\hbar^2} \int^{N t}_0 d\tau_1 \int^{\tau_1}_0 d\tau_2 \eta'(\tau_1)\eta'(\tau_2) \langle \hat{X}(\tau_1)  \hat{X}(\tau_2)\rangle_B} \nonumber \\
 &     e^{\frac{1}{\hbar^2} \int^{N t}_0 d\tau_1 \int^{Nt}_0 d\tau_2 \bar{\eta}'(\tau_1)\eta'(\tau_2) \langle \hat{X}(\tau_1) \hat{X}(\tau_2)\rangle_B}.
\end{align}
Using the decomposition $\eta'(\tau)=\eta^{+}(\tau)+\eta^{-}(\tau)$ and $\bar{\eta}'(\tau)=\eta^{+}(\tau)-\eta^{-}(\tau)$, one can get that
\begin{align}
    \mathcal{G}(\bm{\eta},\bar{\bm{\eta}})=
  &   e^{-\frac{2}{\hbar^2} \int^{N t}_0 d\tau_1 \int^{\tau_1}_0 d\tau_2 \eta^{-}(\tau_1)\eta^{+}(\tau_2) \langle [\hat{X}(\tau_1),  \hat{X}(\tau_2)]\rangle_B} \nonumber \\
 &     e^{-\frac{2}{\hbar^2} \int^{N t}_0 d\tau_1 \int^{Nt}_0 d\tau_2 \eta^{-}(\tau_1)\eta^{-}(\tau_2) \langle\{ \hat{X}(\tau_1), \hat{X}(\tau_2)\}\rangle_B},
\end{align}
where $[\hat{A},\hat{B}]=\hat{A} \hat{B}-\hat{B} \hat{A}$ and  $\{\hat{A},\hat{B}\}=(\hat{A} \hat{B}+\hat{B} \hat{A})/2$.

\section{probability distribution}\label{sec:pd-app}
By using the noise decomposition $\hat{X}(t)=\hat{X}_L(t)+\hat{X}_H(t)$ with the assumption $\langle \hat{X}_H(t) \hat{X}_H(t') \rangle_B=\frac{1}{t_2} \delta(t-t')$, the propagator $\mathcal{G}(\bm{\eta},\bar{\bm{\eta}})$ can be expressed as
\begin{equation}\label{eq:propagator-app}
     \mathcal{G}=  e^{- \sum_{j,l}[2C^{++}_{l,j} \eta^{-}_{l}  \eta^{-}_{j}  + 2i C^{+-}_{l,j}  \eta^{-}_l \eta^{+}_j] -\sum_j (\eta^{-}_j)^2\frac{ t}{t_2}}.
\end{equation} 
To decouple these two terms, we introduce an effective expression of $\eta^{+}$ as $\eta^{+} \equiv (1-(\eta^{-})^2) \sigma$ with $\sigma=\pm 1$. One can check that $\eta^{-}$ and $\sigma$ are independent of each other.  Using this result, the probability distribution $P_{\bm{S}}$ is reduced to 
\begin{equation}\label{eq:Pb-app-I}
    P_{\bm{S}}=\sum_{\bm{\eta}^{-}} \sum_{ \bm{\sigma}} \frac{1}{2^N} \mathcal{G}(\bm{\eta}^{-}, \bm{\sigma}) \prod_{j} \frac{e^{i \eta^{-}_j (\theta+\frac{1-s_j}{2}\pi)}}{4}.
\end{equation}
Because the factor $ \frac{e^{i \eta^{-}_j (\theta+\frac{1-s_j}{2}\pi)}}{4}$ is independent of $\sigma_j$,  the average of $\mathcal{G}$ over  $\{\sigma_j\}$ with probability $p_{\sigma_j=\pm}=1/2$  yields
\begin{align}
    \langle 
   \mathcal{G}\rangle_{\bm{\sigma}}= & e^{-2 \sum_{j, l} C^{++}_{l,j} \eta^{-}_{l} \eta^{-}_{j}  -\sum_j \frac{ ( \eta^{-}_j)^2 t}{t_2}} \nonumber \\
   & \prod_j \cos(2\sum_{l} C^{+-}_{l,j}  \eta^{-}_l \mathcal{P}_j),
\end{align}
where $\mathcal{P}_j=(1-(\eta^{-}_j)^2)$.  Noting further that $|C^{+-}_{l,j} |\ll 1$, one can get that
\begin{align}
   \langle \mathcal{G}\rangle_{\bm{\sigma}} \approx &  e^{- 2\sum_{j, l}C^{++}_{l,j} \eta^{-}_{l}\Delta \eta_{j}-2\sum_j [\sum_{l} C^{+-}_{l,j} \eta^{-}_l ]^2\mathcal{P}_j  -\sum_j \frac{ (\eta^{-}_j)^2 t}{t_2}} \nonumber \\
   \approx &  e^{- 2\sum_{j, l} C^{++}_{l,j}\eta^{-}_{l}\eta^{-}_{j}-2\sum_j [\sum_{l} C^{+-}_{l,j} \eta^{-}_l ]^2  -\sum_j (\eta^{-}_j)^2 \frac{t}{t_2}}.\nonumber
\end{align}
In the last step, we use the fact that $2(\sum_{l} C^{+-}_{l,j} \eta^{-} )^2  \ll t/t_2$.  A further simplification yields that
\begin{equation}
  \langle \mathcal{G}\rangle_{\bm{\sigma}}   \approx e^{- \sum_{l, l'} 2[C^{++}_{l,l'}+\sum_{j}  C^{+-}_{l,j}  C^{+-}_{l',j}]\eta^{-}_{l}\eta^{-}_{l'}  -\sum_l (\eta^{-}_l)^2 \frac{t}{t_2}}.
\end{equation}
To solve the distribution function $P_{\bm{S}}$, we introduce an auxiliary field decomposition of $\langle \mathcal{G}\rangle_{\bm{\sigma}} $, given by
\begin{align*}
e^{-\sum_{l,l'} \eta^{-}_l  D_{l,l'} \eta^{-}_{l'}}=& \frac{1}{\sqrt{\det[\pi D]}} \int  \prod_{j}  d \phi_j  \nonumber \\
& e^{- \sum_{l,l'} \phi_{l} D^{-1}_{l,l'} \phi_{l'} +i2 \sum_{l} \phi_l \eta^{-}_l},
\end{align*}
where $\phi_l$ is denoted as the auxiliary field.  The result is then reduced to 
\begin{equation}
      P_{\bm{S}}= \langle \prod_{j}[\frac{1}{2}+\frac{s_j}{2}\cos (2 \phi_j+ \theta) ] \rangle_{\bm{\phi}},
\end{equation}
where $\langle \rangle_{\bm{\phi}}$ denotes the average over the auxilliary fields$\{\phi_j\}$ with given distribution function $P_{\bm{\phi}}=\frac{e^{- \sum_{l,l'} \phi_{l} D^{-1}_{l,l'} \phi_{l'}} }{\sqrt{\det[\pi D]}}$ with $D_{l,l'}=\delta_{l,l'} \frac{t}{t_2}+2C^{++}_{l,l'}+2\sum_{j}  C^{+-}_{l,j}  C^{+-}_{l',j}$.  

By considering the weak correlation between different auxiliary fields $\{\phi_l\}$ ,  the distribution function $P_{\bm{S}}$ can be  approximated as 
\begin{equation}
          P_{\bm{S}}\approx \prod_{l} P_{s_l} \prod_{l'<l} \frac{P_{s_l,s_l'}}{P_{s_l} P_{s_l'}},
\end{equation}
where   $P_{s_l}=\langle [\frac{1}{2}+\frac{s_l}{2}\cos (2 \phi_l+ \theta) ] \rangle_{\bm{\phi}} $ and $P_{s_l,s_l'}=\langle [\frac{1}{2}+\frac{s_l}{2}\cos (2 \phi_l+ \theta) ][\frac{1}{2}+\frac{s_l'}{2}\cos (2 \phi_l'+ \theta) ] \rangle_{\bm{\phi}}$. By further noting that $\langle \cos(2\phi_l+\theta) \rangle_{\bm{\phi}}=\cos\theta e^{-D_{l,l}}$ and 
$\langle \cos(2\phi_l+\theta)\cos(2\phi_{l'}+\theta) \rangle_{\bm{\phi}}=e^{-D_{l,l}-D_{l',l'}}[\cos^2 \theta \cosh(D_{l,l'}+D_{l',l})+\sin ^2\theta \sinh(D_{l,l'}-D_{l',l})]$, one can get that
\begin{align*}
          P_{\bm{S}}\approx & \prod_{l} P_{s_l} \prod_{l'<l} [1+s_l s_{l'} T_{l,l'}], \\
          T_{l,l'}\equiv & \frac{\cos^2\theta\sinh^2 D_{l,l'}-\sin^2 \theta (1-\sinh{2 D_{l,l'}})}{4 P_{s_l} P_{s_l'}}e^{- D_{l,l}-D_{l',l'}},
\end{align*}
where $D_{l,l'}=D_{l',l}$ is used.  By further ignoring high-order of $C^{+\pm}_{l,l'}$, the final result is then simplified to 
\begin{equation}\label{eq:DF-app}
          P_{\bm{S}}\approx \prod_{l} P_{s_l} \prod_{l'<l} [1-s_l s_{l'} \frac{\sin^2 \theta  C^{++}_{l,l'}}{P_{s_l} P_{s_l'}}e^{- 2t/t_2}],
\end{equation}
where $ P_{s_l} \approx\frac{1}{2}+\frac{s_l}{2}\cos \theta e^{-t/t_2-C^{++}_{l,l}}$.

\bibliography{TU}

\begin{thebibliography}{54}%
\makeatletter
\providecommand \@ifxundefined [1]{%
 \@ifx{#1\undefined}
}%
\providecommand \@ifnum [1]{%
 \ifnum #1\expandafter \@firstoftwo
 \else \expandafter \@secondoftwo
 \fi
}%
\providecommand \@ifx [1]{%
 \ifx #1\expandafter \@firstoftwo
 \else \expandafter \@secondoftwo
 \fi
}%
\providecommand \natexlab [1]{#1}%
\providecommand \enquote  [1]{``#1''}%
\providecommand \bibnamefont  [1]{#1}%
\providecommand \bibfnamefont [1]{#1}%
\providecommand \citenamefont [1]{#1}%
\providecommand \href@noop [0]{\@secondoftwo}%
\providecommand \href [0]{\begingroup \@sanitize@url \@href}%
\providecommand \@href[1]{\@@startlink{#1}\@@href}%
\providecommand \@@href[1]{\endgroup#1\@@endlink}%
\providecommand \@sanitize@url [0]{\catcode `\\12\catcode `\$12\catcode
  `\&12\catcode `\#12\catcode `\^12\catcode `\_12\catcode `\%12\relax}%
\providecommand \@@startlink[1]{}%
\providecommand \@@endlink[0]{}%
\providecommand \url  [0]{\begingroup\@sanitize@url \@url }%
\providecommand \@url [1]{\endgroup\@href {#1}{\urlprefix }}%
\providecommand \urlprefix  [0]{URL }%
\providecommand \Eprint [0]{\href }%
\providecommand \doibase [0]{https://doi.org/}%
\providecommand \selectlanguage [0]{\@gobble}%
\providecommand \bibinfo  [0]{\@secondoftwo}%
\providecommand \bibfield  [0]{\@secondoftwo}%
\providecommand \translation [1]{[#1]}%
\providecommand \BibitemOpen [0]{}%
\providecommand \bibitemStop [0]{}%
\providecommand \bibitemNoStop [0]{.\EOS\space}%
\providecommand \EOS [0]{\spacefactor3000\relax}%
\providecommand \BibitemShut  [1]{\csname bibitem#1\endcsname}%
\let\auto@bib@innerbib\@empty
\bibitem [{\citenamefont {Gross}\ and\ \citenamefont
  {Bloch}(2017)}]{Gross2017}%
  \BibitemOpen
  \bibfield  {author} {\bibinfo {author} {\bibfnamefont {C.}~\bibnamefont
  {Gross}}\ and\ \bibinfo {author} {\bibfnamefont {I.}~\bibnamefont {Bloch}},\
  }\bibfield  {title} {\bibinfo {title} {Quantum simulations with ultracold
  atoms in optical lattices},\ }\href {https://doi.org/10.1126/science.aal3837}
  {\bibfield  {journal} {\bibinfo  {journal} {Science}\ }\textbf {\bibinfo
  {volume} {357}},\ \bibinfo {pages} {995} (\bibinfo {year}
  {2017})}\BibitemShut {NoStop}%
\bibitem [{\citenamefont {Browaeys}\ and\ \citenamefont
  {Lahaye}(2020)}]{Browaeys2020}%
  \BibitemOpen
  \bibfield  {author} {\bibinfo {author} {\bibfnamefont {A.}~\bibnamefont
  {Browaeys}}\ and\ \bibinfo {author} {\bibfnamefont {T.}~\bibnamefont
  {Lahaye}},\ }\bibfield  {title} {\bibinfo {title} {Many-body physics with
  individually controlled rydberg atoms},\ }\href
  {https://doi.org/10.1038/s41567-019-0733-z} {\bibfield  {journal} {\bibinfo
  {journal} {Nat. Phys.}\ }\textbf {\bibinfo {volume} {16}},\ \bibinfo {pages}
  {132} (\bibinfo {year} {2020})}\BibitemShut {NoStop}%
\bibitem [{\citenamefont {Ebadi}\ \emph {et~al.}(2021)\citenamefont {Ebadi},
  \citenamefont {Wang}, \citenamefont {Levine}, \citenamefont {Keesling},
  \citenamefont {Semeghini}, \citenamefont {Omran}, \citenamefont {Bluvstein},
  \citenamefont {Samajdar}, \citenamefont {Pichler}, \citenamefont {Ho},
  \citenamefont {Choi}, \citenamefont {Sachdev}, \citenamefont {Greiner},
  \citenamefont {Vuletić},\ and\ \citenamefont {Lukin}}]{Ebadi2021}%
  \BibitemOpen
  \bibfield  {author} {\bibinfo {author} {\bibfnamefont {S.}~\bibnamefont
  {Ebadi}}, \bibinfo {author} {\bibfnamefont {T.~T.}\ \bibnamefont {Wang}},
  \bibinfo {author} {\bibfnamefont {H.}~\bibnamefont {Levine}}, \bibinfo
  {author} {\bibfnamefont {A.}~\bibnamefont {Keesling}}, \bibinfo {author}
  {\bibfnamefont {G.}~\bibnamefont {Semeghini}}, \bibinfo {author}
  {\bibfnamefont {A.}~\bibnamefont {Omran}}, \bibinfo {author} {\bibfnamefont
  {D.}~\bibnamefont {Bluvstein}}, \bibinfo {author} {\bibfnamefont
  {R.}~\bibnamefont {Samajdar}}, \bibinfo {author} {\bibfnamefont
  {H.}~\bibnamefont {Pichler}}, \bibinfo {author} {\bibfnamefont {W.~W.}\
  \bibnamefont {Ho}}, \bibinfo {author} {\bibfnamefont {S.}~\bibnamefont
  {Choi}}, \bibinfo {author} {\bibfnamefont {S.}~\bibnamefont {Sachdev}},
  \bibinfo {author} {\bibfnamefont {M.}~\bibnamefont {Greiner}}, \bibinfo
  {author} {\bibfnamefont {V.}~\bibnamefont {Vuletić}},\ and\ \bibinfo
  {author} {\bibfnamefont {M.~D.}\ \bibnamefont {Lukin}},\ }\bibfield  {title}
  {\bibinfo {title} {Quantum phases of matter on a 256-atom programmable
  quantum simulator},\ }\href {https://doi.org/10.1038/s41586-021-03582-4}
  {\bibfield  {journal} {\bibinfo  {journal} {Nature}\ }\textbf {\bibinfo
  {volume} {595}},\ \bibinfo {pages} {227} (\bibinfo {year}
  {2021})}\BibitemShut {NoStop}%
\bibitem [{\citenamefont {Morgado}\ and\ \citenamefont
  {Whitlock}(2021)}]{Morgado2021}%
  \BibitemOpen
  \bibfield  {author} {\bibinfo {author} {\bibfnamefont {M.}~\bibnamefont
  {Morgado}}\ and\ \bibinfo {author} {\bibfnamefont {S.}~\bibnamefont
  {Whitlock}},\ }\bibfield  {title} {\bibinfo {title} {{Quantum simulation and
  computing with Rydberg-interacting qubits}},\ }\href
  {https://doi.org/10.1116/5.0036562} {\bibfield  {journal} {\bibinfo
  {journal} {AVS Quantum Science}\ }\textbf {\bibinfo {volume} {3}},\ \bibinfo
  {pages} {023501} (\bibinfo {year} {2021})}\BibitemShut {NoStop}%
\bibitem [{\citenamefont {Leanhardt}\ \emph {et~al.}(2003)\citenamefont
  {Leanhardt}, \citenamefont {Pasquini}, \citenamefont {Saba}, \citenamefont
  {Schirotzek}, \citenamefont {Shin}, \citenamefont {Kielpinski}, \citenamefont
  {Pritchard},\ and\ \citenamefont {Ketterle}}]{Leanhardt2003}%
  \BibitemOpen
  \bibfield  {author} {\bibinfo {author} {\bibfnamefont {A.~E.}\ \bibnamefont
  {Leanhardt}}, \bibinfo {author} {\bibfnamefont {T.~A.}\ \bibnamefont
  {Pasquini}}, \bibinfo {author} {\bibfnamefont {M.}~\bibnamefont {Saba}},
  \bibinfo {author} {\bibfnamefont {A.}~\bibnamefont {Schirotzek}}, \bibinfo
  {author} {\bibfnamefont {Y.}~\bibnamefont {Shin}}, \bibinfo {author}
  {\bibfnamefont {D.}~\bibnamefont {Kielpinski}}, \bibinfo {author}
  {\bibfnamefont {D.~E.}\ \bibnamefont {Pritchard}},\ and\ \bibinfo {author}
  {\bibfnamefont {W.}~\bibnamefont {Ketterle}},\ }\bibfield  {title} {\bibinfo
  {title} {Cooling bose-einstein condensates below 500 picokelvin},\ }\href
  {https://doi.org/10.1126/science.1088827} {\bibfield  {journal} {\bibinfo
  {journal} {Science}\ }\textbf {\bibinfo {volume} {301}},\ \bibinfo {pages}
  {1513} (\bibinfo {year} {2003})}\BibitemShut {NoStop}%
\bibitem [{\citenamefont {Gati}\ \emph {et~al.}(2006)\citenamefont {Gati},
  \citenamefont {Hemmerling}, \citenamefont {F\"olling}, \citenamefont
  {Albiez},\ and\ \citenamefont {Oberthaler}}]{Gati2006}%
  \BibitemOpen
  \bibfield  {author} {\bibinfo {author} {\bibfnamefont {R.}~\bibnamefont
  {Gati}}, \bibinfo {author} {\bibfnamefont {B.}~\bibnamefont {Hemmerling}},
  \bibinfo {author} {\bibfnamefont {J.}~\bibnamefont {F\"olling}}, \bibinfo
  {author} {\bibfnamefont {M.}~\bibnamefont {Albiez}},\ and\ \bibinfo {author}
  {\bibfnamefont {M.~K.}\ \bibnamefont {Oberthaler}},\ }\bibfield  {title}
  {\bibinfo {title} {Noise thermometry with two weakly coupled bose-einstein
  condensates},\ }\href {https://doi.org/10.1103/PhysRevLett.96.130404}
  {\bibfield  {journal} {\bibinfo  {journal} {Phys. Rev. Lett.}\ }\textbf
  {\bibinfo {volume} {96}},\ \bibinfo {pages} {130404} (\bibinfo {year}
  {2006})}\BibitemShut {NoStop}%
\bibitem [{\citenamefont {Correa}\ \emph {et~al.}(2017)\citenamefont {Correa},
  \citenamefont {Perarnau-Llobet}, \citenamefont {Hovhannisyan}, \citenamefont
  {Hern\'andez-Santana}, \citenamefont {Mehboudi},\ and\ \citenamefont
  {Sanpera}}]{Correa2017}%
  \BibitemOpen
  \bibfield  {author} {\bibinfo {author} {\bibfnamefont {L.~A.}\ \bibnamefont
  {Correa}}, \bibinfo {author} {\bibfnamefont {M.}~\bibnamefont
  {Perarnau-Llobet}}, \bibinfo {author} {\bibfnamefont {K.~V.}\ \bibnamefont
  {Hovhannisyan}}, \bibinfo {author} {\bibfnamefont {S.}~\bibnamefont
  {Hern\'andez-Santana}}, \bibinfo {author} {\bibfnamefont {M.}~\bibnamefont
  {Mehboudi}},\ and\ \bibinfo {author} {\bibfnamefont {A.}~\bibnamefont
  {Sanpera}},\ }\bibfield  {title} {\bibinfo {title} {Enhancement of
  low-temperature thermometry by strong coupling},\ }\href
  {https://doi.org/10.1103/PhysRevA.96.062103} {\bibfield  {journal} {\bibinfo
  {journal} {Phys. Rev. A}\ }\textbf {\bibinfo {volume} {96}},\ \bibinfo
  {pages} {062103} (\bibinfo {year} {2017})}\BibitemShut {NoStop}%
\bibitem [{\citenamefont {Mehboudi}\ \emph
  {et~al.}(2019{\natexlab{a}})\citenamefont {Mehboudi}, \citenamefont {Lampo},
  \citenamefont {Charalambous}, \citenamefont {Correa}, \citenamefont
  {Garc\'{\i}a-March},\ and\ \citenamefont {Lewenstein}}]{Mehboudi2019}%
  \BibitemOpen
  \bibfield  {author} {\bibinfo {author} {\bibfnamefont {M.}~\bibnamefont
  {Mehboudi}}, \bibinfo {author} {\bibfnamefont {A.}~\bibnamefont {Lampo}},
  \bibinfo {author} {\bibfnamefont {C.}~\bibnamefont {Charalambous}}, \bibinfo
  {author} {\bibfnamefont {L.~A.}\ \bibnamefont {Correa}}, \bibinfo {author}
  {\bibfnamefont {M.~A.}\ \bibnamefont {Garc\'{\i}a-March}},\ and\ \bibinfo
  {author} {\bibfnamefont {M.}~\bibnamefont {Lewenstein}},\ }\bibfield  {title}
  {\bibinfo {title} {Using polarons for sub-nk quantum nondemolition
  thermometry in a bose-einstein condensate},\ }\href
  {https://doi.org/10.1103/PhysRevLett.122.030403} {\bibfield  {journal}
  {\bibinfo  {journal} {Phys. Rev. Lett.}\ }\textbf {\bibinfo {volume} {122}},\
  \bibinfo {pages} {030403} (\bibinfo {year} {2019}{\natexlab{a}})}\BibitemShut
  {NoStop}%
\bibitem [{\citenamefont {Mitchison}\ \emph {et~al.}(2020)\citenamefont
  {Mitchison}, \citenamefont {Fogarty}, \citenamefont {Guarnieri},
  \citenamefont {Campbell}, \citenamefont {Busch},\ and\ \citenamefont
  {Goold}}]{Mitchison2020}%
  \BibitemOpen
  \bibfield  {author} {\bibinfo {author} {\bibfnamefont {M.~T.}\ \bibnamefont
  {Mitchison}}, \bibinfo {author} {\bibfnamefont {T.}~\bibnamefont {Fogarty}},
  \bibinfo {author} {\bibfnamefont {G.}~\bibnamefont {Guarnieri}}, \bibinfo
  {author} {\bibfnamefont {S.}~\bibnamefont {Campbell}}, \bibinfo {author}
  {\bibfnamefont {T.}~\bibnamefont {Busch}},\ and\ \bibinfo {author}
  {\bibfnamefont {J.}~\bibnamefont {Goold}},\ }\bibfield  {title} {\bibinfo
  {title} {In situ thermometry of a cold fermi gas via dephasing impurities},\
  }\href {https://doi.org/10.1103/PhysRevLett.125.080402} {\bibfield  {journal}
  {\bibinfo  {journal} {Phys. Rev. Lett.}\ }\textbf {\bibinfo {volume} {125}},\
  \bibinfo {pages} {080402} (\bibinfo {year} {2020})}\BibitemShut {NoStop}%
\bibitem [{\citenamefont {Bouton}\ \emph {et~al.}(2020)\citenamefont {Bouton},
  \citenamefont {Nettersheim}, \citenamefont {Adam}, \citenamefont {Schmidt},
  \citenamefont {Mayer}, \citenamefont {Lausch}, \citenamefont {Tiemann},\ and\
  \citenamefont {Widera}}]{Bouton2020}%
  \BibitemOpen
  \bibfield  {author} {\bibinfo {author} {\bibfnamefont {Q.}~\bibnamefont
  {Bouton}}, \bibinfo {author} {\bibfnamefont {J.}~\bibnamefont {Nettersheim}},
  \bibinfo {author} {\bibfnamefont {D.}~\bibnamefont {Adam}}, \bibinfo {author}
  {\bibfnamefont {F.}~\bibnamefont {Schmidt}}, \bibinfo {author} {\bibfnamefont
  {D.}~\bibnamefont {Mayer}}, \bibinfo {author} {\bibfnamefont
  {T.}~\bibnamefont {Lausch}}, \bibinfo {author} {\bibfnamefont
  {E.}~\bibnamefont {Tiemann}},\ and\ \bibinfo {author} {\bibfnamefont
  {A.}~\bibnamefont {Widera}},\ }\bibfield  {title} {\bibinfo {title}
  {Single-atom quantum probes for ultracold gases boosted by nonequilibrium
  spin dynamics},\ }\href {https://doi.org/10.1103/PhysRevX.10.011018}
  {\bibfield  {journal} {\bibinfo  {journal} {Phys. Rev. X}\ }\textbf {\bibinfo
  {volume} {10}},\ \bibinfo {pages} {011018} (\bibinfo {year}
  {2020})}\BibitemShut {NoStop}%
\bibitem [{\citenamefont {Adam}\ \emph {et~al.}(2022)\citenamefont {Adam},
  \citenamefont {Bouton}, \citenamefont {Nettersheim}, \citenamefont
  {Burgardt},\ and\ \citenamefont {Widera}}]{Adam2022}%
  \BibitemOpen
  \bibfield  {author} {\bibinfo {author} {\bibfnamefont {D.}~\bibnamefont
  {Adam}}, \bibinfo {author} {\bibfnamefont {Q.}~\bibnamefont {Bouton}},
  \bibinfo {author} {\bibfnamefont {J.}~\bibnamefont {Nettersheim}}, \bibinfo
  {author} {\bibfnamefont {S.}~\bibnamefont {Burgardt}},\ and\ \bibinfo
  {author} {\bibfnamefont {A.}~\bibnamefont {Widera}},\ }\bibfield  {title}
  {\bibinfo {title} {Coherent and dephasing spectroscopy for single-impurity
  probing of an ultracold bath},\ }\href
  {https://doi.org/10.1103/PhysRevLett.129.120404} {\bibfield  {journal}
  {\bibinfo  {journal} {Phys. Rev. Lett.}\ }\textbf {\bibinfo {volume} {129}},\
  \bibinfo {pages} {120404} (\bibinfo {year} {2022})}\BibitemShut {NoStop}%
\bibitem [{\citenamefont {Stace}(2010)}]{Stace2010}%
  \BibitemOpen
  \bibfield  {author} {\bibinfo {author} {\bibfnamefont {T.~M.}\ \bibnamefont
  {Stace}},\ }\bibfield  {title} {\bibinfo {title} {Quantum limits of
  thermometry},\ }\href {https://doi.org/10.1103/PhysRevA.82.011611} {\bibfield
   {journal} {\bibinfo  {journal} {Phys. Rev. A}\ }\textbf {\bibinfo {volume}
  {82}},\ \bibinfo {pages} {011611} (\bibinfo {year} {2010})}\BibitemShut
  {NoStop}%
\bibitem [{\citenamefont {Johnson}\ \emph {et~al.}(2016)\citenamefont
  {Johnson}, \citenamefont {Cosco}, \citenamefont {Mitchison}, \citenamefont
  {Jaksch},\ and\ \citenamefont {Clark}}]{Johnson2016}%
  \BibitemOpen
  \bibfield  {author} {\bibinfo {author} {\bibfnamefont {T.~H.}\ \bibnamefont
  {Johnson}}, \bibinfo {author} {\bibfnamefont {F.}~\bibnamefont {Cosco}},
  \bibinfo {author} {\bibfnamefont {M.~T.}\ \bibnamefont {Mitchison}}, \bibinfo
  {author} {\bibfnamefont {D.}~\bibnamefont {Jaksch}},\ and\ \bibinfo {author}
  {\bibfnamefont {S.~R.}\ \bibnamefont {Clark}},\ }\bibfield  {title} {\bibinfo
  {title} {Thermometry of ultracold atoms via nonequilibrium work
  distributions},\ }\href {https://doi.org/10.1103/PhysRevA.93.053619}
  {\bibfield  {journal} {\bibinfo  {journal} {Phys. Rev. A}\ }\textbf {\bibinfo
  {volume} {93}},\ \bibinfo {pages} {053619} (\bibinfo {year}
  {2016})}\BibitemShut {NoStop}%
\bibitem [{\citenamefont {Razavian}\ \emph {et~al.}(2019)\citenamefont
  {Razavian}, \citenamefont {Benedetti}, \citenamefont {Bina}, \citenamefont
  {Akbari-Kourbolagh},\ and\ \citenamefont {Paris}}]{Razavian2019}%
  \BibitemOpen
  \bibfield  {author} {\bibinfo {author} {\bibfnamefont {S.}~\bibnamefont
  {Razavian}}, \bibinfo {author} {\bibfnamefont {C.}~\bibnamefont {Benedetti}},
  \bibinfo {author} {\bibfnamefont {M.}~\bibnamefont {Bina}}, \bibinfo {author}
  {\bibfnamefont {Y.}~\bibnamefont {Akbari-Kourbolagh}},\ and\ \bibinfo
  {author} {\bibfnamefont {M.~G.~A.}\ \bibnamefont {Paris}},\ }\bibfield
  {title} {\bibinfo {title} {Quantum thermometry by single-qubit dephasing},\
  }\href {https://doi.org/10.1140/epjp/i2019-12708-9} {\bibfield  {journal}
  {\bibinfo  {journal} {Eur. Phys. J. Plus}\ }\textbf {\bibinfo {volume}
  {134}},\ \bibinfo {pages} {284} (\bibinfo {year} {2019})}\BibitemShut
  {NoStop}%
\bibitem [{\citenamefont {Yuan}\ \emph {et~al.}(2023)\citenamefont {Yuan},
  \citenamefont {Zhang}, \citenamefont {Song}, \citenamefont {Tang},
  \citenamefont {Wang},\ and\ \citenamefont {Kuang}}]{Yuan2023}%
  \BibitemOpen
  \bibfield  {author} {\bibinfo {author} {\bibfnamefont {J.-B.}\ \bibnamefont
  {Yuan}}, \bibinfo {author} {\bibfnamefont {B.}~\bibnamefont {Zhang}},
  \bibinfo {author} {\bibfnamefont {Y.-J.}\ \bibnamefont {Song}}, \bibinfo
  {author} {\bibfnamefont {S.-Q.}\ \bibnamefont {Tang}}, \bibinfo {author}
  {\bibfnamefont {X.-W.}\ \bibnamefont {Wang}},\ and\ \bibinfo {author}
  {\bibfnamefont {L.-M.}\ \bibnamefont {Kuang}},\ }\bibfield  {title} {\bibinfo
  {title} {Quantum sensing of temperature close to absolute zero in a
  bose-einstein condensate},\ }\href
  {https://doi.org/10.1103/PhysRevA.107.063317} {\bibfield  {journal} {\bibinfo
   {journal} {Phys. Rev. A}\ }\textbf {\bibinfo {volume} {107}},\ \bibinfo
  {pages} {063317} (\bibinfo {year} {2023})}\BibitemShut {NoStop}%
\bibitem [{\citenamefont {Mehboudi}\ \emph
  {et~al.}(2019{\natexlab{b}})\citenamefont {Mehboudi}, \citenamefont
  {Sanpera},\ and\ \citenamefont {Correa}}]{Mehboudi2019b}%
  \BibitemOpen
  \bibfield  {author} {\bibinfo {author} {\bibfnamefont {M.}~\bibnamefont
  {Mehboudi}}, \bibinfo {author} {\bibfnamefont {A.}~\bibnamefont {Sanpera}},\
  and\ \bibinfo {author} {\bibfnamefont {L.~A.}\ \bibnamefont {Correa}},\
  }\bibfield  {title} {\bibinfo {title} {Thermometry in the quantum regime:
  recent theoretical progress},\ }\href
  {https://doi.org/10.1088/1751-8121/ab2828} {\bibfield  {journal} {\bibinfo
  {journal} {J. Phys. A}\ }\textbf {\bibinfo {volume} {52}},\ \bibinfo {pages}
  {303001} (\bibinfo {year} {2019}{\natexlab{b}})}\BibitemShut {NoStop}%
\bibitem [{\citenamefont {Potts}\ \emph {et~al.}(2019)\citenamefont {Potts},
  \citenamefont {Brask},\ and\ \citenamefont {Brunner}}]{Potts2019}%
  \BibitemOpen
  \bibfield  {author} {\bibinfo {author} {\bibfnamefont {P.~P.}\ \bibnamefont
  {Potts}}, \bibinfo {author} {\bibfnamefont {J.~B.}\ \bibnamefont {Brask}},\
  and\ \bibinfo {author} {\bibfnamefont {N.}~\bibnamefont {Brunner}},\
  }\bibfield  {title} {\bibinfo {title} {Fundamental limits on low-temperature
  quantum thermometry with finite resolution},\ }\href
  {https://doi.org/10.22331/q-2019-07-09-161} {\bibfield  {journal} {\bibinfo
  {journal} {{Quantum}}\ }\textbf {\bibinfo {volume} {3}},\ \bibinfo {pages}
  {161} (\bibinfo {year} {2019})}\BibitemShut {NoStop}%
\bibitem [{\citenamefont {J\o{}rgensen}\ \emph {et~al.}(2020)\citenamefont
  {J\o{}rgensen}, \citenamefont {Potts}, \citenamefont {Paris},\ and\
  \citenamefont {Brask}}]{Jorgensen2020}%
  \BibitemOpen
  \bibfield  {author} {\bibinfo {author} {\bibfnamefont {M.~R.}\ \bibnamefont
  {J\o{}rgensen}}, \bibinfo {author} {\bibfnamefont {P.~P.}\ \bibnamefont
  {Potts}}, \bibinfo {author} {\bibfnamefont {M.~G.~A.}\ \bibnamefont
  {Paris}},\ and\ \bibinfo {author} {\bibfnamefont {J.~B.}\ \bibnamefont
  {Brask}},\ }\bibfield  {title} {\bibinfo {title} {Tight bound on
  finite-resolution quantum thermometry at low temperatures},\ }\href
  {https://doi.org/10.1103/PhysRevResearch.2.033394} {\bibfield  {journal}
  {\bibinfo  {journal} {Phys. Rev. Res.}\ }\textbf {\bibinfo {volume} {2}},\
  \bibinfo {pages} {033394} (\bibinfo {year} {2020})}\BibitemShut {NoStop}%
\bibitem [{\citenamefont {Mihailescu}\ \emph {et~al.}(2023)\citenamefont
  {Mihailescu}, \citenamefont {Campbell},\ and\ \citenamefont
  {Mitchell}}]{Mihailescu2023}%
  \BibitemOpen
  \bibfield  {author} {\bibinfo {author} {\bibfnamefont {G.}~\bibnamefont
  {Mihailescu}}, \bibinfo {author} {\bibfnamefont {S.}~\bibnamefont
  {Campbell}},\ and\ \bibinfo {author} {\bibfnamefont {A.~K.}\ \bibnamefont
  {Mitchell}},\ }\bibfield  {title} {\bibinfo {title} {Thermometry of strongly
  correlated fermionic quantum systems using impurity probes},\ }\href
  {https://doi.org/10.1103/PhysRevA.107.042614} {\bibfield  {journal} {\bibinfo
   {journal} {Phys. Rev. A}\ }\textbf {\bibinfo {volume} {107}},\ \bibinfo
  {pages} {042614} (\bibinfo {year} {2023})}\BibitemShut {NoStop}%
\bibitem [{\citenamefont {Brenes}\ and\ \citenamefont
  {Segal}(2023)}]{Brenes2023}%
  \BibitemOpen
  \bibfield  {author} {\bibinfo {author} {\bibfnamefont {M.}~\bibnamefont
  {Brenes}}\ and\ \bibinfo {author} {\bibfnamefont {D.}~\bibnamefont {Segal}},\
  }\bibfield  {title} {\bibinfo {title} {Multispin probes for thermometry in
  the strong-coupling regime},\ }\href
  {https://doi.org/10.1103/PhysRevA.108.032220} {\bibfield  {journal} {\bibinfo
   {journal} {Phys. Rev. A}\ }\textbf {\bibinfo {volume} {108}},\ \bibinfo
  {pages} {032220} (\bibinfo {year} {2023})}\BibitemShut {NoStop}%
\bibitem [{\citenamefont {Seah}\ \emph {et~al.}(2019)\citenamefont {Seah},
  \citenamefont {Nimmrichter}, \citenamefont {Grimmer}, \citenamefont {Santos},
  \citenamefont {Scarani},\ and\ \citenamefont {Landi}}]{Seah2019}%
  \BibitemOpen
  \bibfield  {author} {\bibinfo {author} {\bibfnamefont {S.}~\bibnamefont
  {Seah}}, \bibinfo {author} {\bibfnamefont {S.}~\bibnamefont {Nimmrichter}},
  \bibinfo {author} {\bibfnamefont {D.}~\bibnamefont {Grimmer}}, \bibinfo
  {author} {\bibfnamefont {J.~P.}\ \bibnamefont {Santos}}, \bibinfo {author}
  {\bibfnamefont {V.}~\bibnamefont {Scarani}},\ and\ \bibinfo {author}
  {\bibfnamefont {G.~T.}\ \bibnamefont {Landi}},\ }\bibfield  {title} {\bibinfo
  {title} {Collisional quantum thermometry},\ }\href
  {https://doi.org/10.1103/PhysRevLett.123.180602} {\bibfield  {journal}
  {\bibinfo  {journal} {Phys. Rev. Lett.}\ }\textbf {\bibinfo {volume} {123}},\
  \bibinfo {pages} {180602} (\bibinfo {year} {2019})}\BibitemShut {NoStop}%
\bibitem [{\citenamefont {Alves}\ and\ \citenamefont
  {Landi}(2022)}]{Alves2022}%
  \BibitemOpen
  \bibfield  {author} {\bibinfo {author} {\bibfnamefont {G.~O.}\ \bibnamefont
  {Alves}}\ and\ \bibinfo {author} {\bibfnamefont {G.~T.}\ \bibnamefont
  {Landi}},\ }\bibfield  {title} {\bibinfo {title} {Bayesian estimation for
  collisional thermometry},\ }\href
  {https://doi.org/10.1103/PhysRevA.105.012212} {\bibfield  {journal} {\bibinfo
   {journal} {Phys. Rev. A}\ }\textbf {\bibinfo {volume} {105}},\ \bibinfo
  {pages} {012212} (\bibinfo {year} {2022})}\BibitemShut {NoStop}%
\bibitem [{\citenamefont {Hovhannisyan}\ and\ \citenamefont
  {Correa}(2018)}]{Hovhannisyan2018}%
  \BibitemOpen
  \bibfield  {author} {\bibinfo {author} {\bibfnamefont {K.~V.}\ \bibnamefont
  {Hovhannisyan}}\ and\ \bibinfo {author} {\bibfnamefont {L.~A.}\ \bibnamefont
  {Correa}},\ }\bibfield  {title} {\bibinfo {title} {Measuring the temperature
  of cold many-body quantum systems},\ }\href
  {https://doi.org/10.1103/PhysRevB.98.045101} {\bibfield  {journal} {\bibinfo
  {journal} {Phys. Rev. B}\ }\textbf {\bibinfo {volume} {98}},\ \bibinfo
  {pages} {045101} (\bibinfo {year} {2018})}\BibitemShut {NoStop}%
\bibitem [{\citenamefont {Mirkhalaf}\ \emph {et~al.}(2021)\citenamefont
  {Mirkhalaf}, \citenamefont {Benedicto~Orenes}, \citenamefont {Mitchell},\
  and\ \citenamefont {Witkowska}}]{Mirkhalaf2021}%
  \BibitemOpen
  \bibfield  {author} {\bibinfo {author} {\bibfnamefont {S.~S.}\ \bibnamefont
  {Mirkhalaf}}, \bibinfo {author} {\bibfnamefont {D.}~\bibnamefont
  {Benedicto~Orenes}}, \bibinfo {author} {\bibfnamefont {M.~W.}\ \bibnamefont
  {Mitchell}},\ and\ \bibinfo {author} {\bibfnamefont {E.}~\bibnamefont
  {Witkowska}},\ }\bibfield  {title} {\bibinfo {title} {Criticality-enhanced
  quantum sensing in ferromagnetic bose-einstein condensates: Role of readout
  measurement and detection noise},\ }\href
  {https://doi.org/10.1103/PhysRevA.103.023317} {\bibfield  {journal} {\bibinfo
   {journal} {Phys. Rev. A}\ }\textbf {\bibinfo {volume} {103}},\ \bibinfo
  {pages} {023317} (\bibinfo {year} {2021})}\BibitemShut {NoStop}%
\bibitem [{\citenamefont {Aybar}\ \emph {et~al.}(2022)\citenamefont {Aybar},
  \citenamefont {Niezgoda}, \citenamefont {Mirkhalaf}, \citenamefont
  {Mitchell}, \citenamefont {Benedicto~Orenes},\ and\ \citenamefont
  {Witkowska}}]{Aybar2022}%
  \BibitemOpen
  \bibfield  {author} {\bibinfo {author} {\bibfnamefont {E.}~\bibnamefont
  {Aybar}}, \bibinfo {author} {\bibfnamefont {A.}~\bibnamefont {Niezgoda}},
  \bibinfo {author} {\bibfnamefont {S.~S.}\ \bibnamefont {Mirkhalaf}}, \bibinfo
  {author} {\bibfnamefont {M.~W.}\ \bibnamefont {Mitchell}}, \bibinfo {author}
  {\bibfnamefont {D.}~\bibnamefont {Benedicto~Orenes}},\ and\ \bibinfo {author}
  {\bibfnamefont {E.}~\bibnamefont {Witkowska}},\ }\bibfield  {title} {\bibinfo
  {title} {Critical quantum thermometry and its feasibility in spin systems},\
  }\href {https://doi.org/10.22331/q-2022-09-19-808} {\bibfield  {journal}
  {\bibinfo  {journal} {{Quantum}}\ }\textbf {\bibinfo {volume} {6}},\ \bibinfo
  {pages} {808} (\bibinfo {year} {2022})}\BibitemShut {NoStop}%
\bibitem [{\citenamefont {Zhang}\ \emph {et~al.}(2022)\citenamefont {Zhang},
  \citenamefont {Chen}, \citenamefont {Bai}, \citenamefont {Wu},\ and\
  \citenamefont {An}}]{Zhang2022}%
  \BibitemOpen
  \bibfield  {author} {\bibinfo {author} {\bibfnamefont {N.}~\bibnamefont
  {Zhang}}, \bibinfo {author} {\bibfnamefont {C.}~\bibnamefont {Chen}},
  \bibinfo {author} {\bibfnamefont {S.-Y.}\ \bibnamefont {Bai}}, \bibinfo
  {author} {\bibfnamefont {W.}~\bibnamefont {Wu}},\ and\ \bibinfo {author}
  {\bibfnamefont {J.-H.}\ \bibnamefont {An}},\ }\bibfield  {title} {\bibinfo
  {title} {Non-markovian quantum thermometry},\ }\href
  {https://doi.org/10.1103/PhysRevApplied.17.034073} {\bibfield  {journal}
  {\bibinfo  {journal} {Phys. Rev. Appl.}\ }\textbf {\bibinfo {volume} {17}},\
  \bibinfo {pages} {034073} (\bibinfo {year} {2022})}\BibitemShut {NoStop}%
\bibitem [{\citenamefont {Zhang}\ and\ \citenamefont {Wu}(2021)}]{Zhang2021}%
  \BibitemOpen
  \bibfield  {author} {\bibinfo {author} {\bibfnamefont {Z.-Z.}\ \bibnamefont
  {Zhang}}\ and\ \bibinfo {author} {\bibfnamefont {W.}~\bibnamefont {Wu}},\
  }\bibfield  {title} {\bibinfo {title} {Non-markovian temperature sensing},\
  }\href {https://doi.org/10.1103/PhysRevResearch.3.043039} {\bibfield
  {journal} {\bibinfo  {journal} {Phys. Rev. Res.}\ }\textbf {\bibinfo {volume}
  {3}},\ \bibinfo {pages} {043039} (\bibinfo {year} {2021})}\BibitemShut
  {NoStop}%
\bibitem [{\citenamefont {Xu}\ \emph {et~al.}(2023)\citenamefont {Xu},
  \citenamefont {Yuan}, \citenamefont {Tang}, \citenamefont {Wu}, \citenamefont
  {Tan},\ and\ \citenamefont {Kuang}}]{Xu2023}%
  \BibitemOpen
  \bibfield  {author} {\bibinfo {author} {\bibfnamefont {L.}~\bibnamefont
  {Xu}}, \bibinfo {author} {\bibfnamefont {J.-B.}\ \bibnamefont {Yuan}},
  \bibinfo {author} {\bibfnamefont {S.-Q.}\ \bibnamefont {Tang}}, \bibinfo
  {author} {\bibfnamefont {W.}~\bibnamefont {Wu}}, \bibinfo {author}
  {\bibfnamefont {Q.-S.}\ \bibnamefont {Tan}},\ and\ \bibinfo {author}
  {\bibfnamefont {L.-M.}\ \bibnamefont {Kuang}},\ }\bibfield  {title} {\bibinfo
  {title} {Non-markovian enhanced temperature sensing in a dipolar
  bose-einstein condensate},\ }\href
  {https://doi.org/10.1103/PhysRevA.108.022608} {\bibfield  {journal} {\bibinfo
   {journal} {Phys. Rev. A}\ }\textbf {\bibinfo {volume} {108}},\ \bibinfo
  {pages} {022608} (\bibinfo {year} {2023})}\BibitemShut {NoStop}%
\bibitem [{\citenamefont {Zhang}\ \emph {et~al.}(2024)\citenamefont {Zhang},
  \citenamefont {Bai},\ and\ \citenamefont {Chen}}]{Zhang2023}%
  \BibitemOpen
  \bibfield  {author} {\bibinfo {author} {\bibfnamefont {N.}~\bibnamefont
  {Zhang}}, \bibinfo {author} {\bibfnamefont {S.-Y.}\ \bibnamefont {Bai}},\
  and\ \bibinfo {author} {\bibfnamefont {C.}~\bibnamefont {Chen}},\ }\bibfield
  {title} {\bibinfo {title} {Temperature-heat uncertainty relation in
  nonequilibrium quantum thermometry},\ }\href
  {https://doi.org/10.1103/PhysRevA.110.012211} {\bibfield  {journal} {\bibinfo
   {journal} {Phys. Rev. A}\ }\textbf {\bibinfo {volume} {110}},\ \bibinfo
  {pages} {012211} (\bibinfo {year} {2024})}\BibitemShut {NoStop}%
\bibitem [{\citenamefont {Planella}\ \emph {et~al.}(2022)\citenamefont
  {Planella}, \citenamefont {Cenni}, \citenamefont {Ac\'{\i}n},\ and\
  \citenamefont {Mehboudi}}]{Planella2022}%
  \BibitemOpen
  \bibfield  {author} {\bibinfo {author} {\bibfnamefont {G.}~\bibnamefont
  {Planella}}, \bibinfo {author} {\bibfnamefont {M.~F.~B.}\ \bibnamefont
  {Cenni}}, \bibinfo {author} {\bibfnamefont {A.}~\bibnamefont {Ac\'{\i}n}},\
  and\ \bibinfo {author} {\bibfnamefont {M.}~\bibnamefont {Mehboudi}},\
  }\bibfield  {title} {\bibinfo {title} {Bath-induced correlations enhance
  thermometry precision at low temperatures},\ }\href
  {https://doi.org/10.1103/PhysRevLett.128.040502} {\bibfield  {journal}
  {\bibinfo  {journal} {Phys. Rev. Lett.}\ }\textbf {\bibinfo {volume} {128}},\
  \bibinfo {pages} {040502} (\bibinfo {year} {2022})}\BibitemShut {NoStop}%
\bibitem [{\citenamefont {Brattegard}\ and\ \citenamefont
  {Mitchison}(2024)}]{Brattegard2024}%
  \BibitemOpen
  \bibfield  {author} {\bibinfo {author} {\bibfnamefont {S.}~\bibnamefont
  {Brattegard}}\ and\ \bibinfo {author} {\bibfnamefont {M.~T.}\ \bibnamefont
  {Mitchison}},\ }\bibfield  {title} {\bibinfo {title} {Thermometry by
  correlated dephasing of impurities in a one-dimensional fermi gas},\ }\href
  {https://doi.org/10.1103/PhysRevA.109.023309} {\bibfield  {journal} {\bibinfo
   {journal} {Phys. Rev. A}\ }\textbf {\bibinfo {volume} {109}},\ \bibinfo
  {pages} {023309} (\bibinfo {year} {2024})}\BibitemShut {NoStop}%
\bibitem [{\citenamefont {Zhang}\ and\ \citenamefont {Chen}()}]{Zhang2024}%
  \BibitemOpen
  \bibfield  {author} {\bibinfo {author} {\bibfnamefont {N.}~\bibnamefont
  {Zhang}}\ and\ \bibinfo {author} {\bibfnamefont {C.}~\bibnamefont {Chen}},\
  }\href@noop {} {\bibinfo {title} {Achieving heisenberg scaling in
  low-temperature quantum thermometry}},\ \Eprint
  {https://arxiv.org/abs/2407.05762} {arXiv:2407.05762} \BibitemShut {NoStop}%
\bibitem [{\citenamefont {Burgarth}\ \emph {et~al.}(2015)\citenamefont
  {Burgarth}, \citenamefont {Giovannetti}, \citenamefont {Kato},\ and\
  \citenamefont {Yuasa}}]{Burgarth2015}%
  \BibitemOpen
  \bibfield  {author} {\bibinfo {author} {\bibfnamefont {D.}~\bibnamefont
  {Burgarth}}, \bibinfo {author} {\bibfnamefont {V.}~\bibnamefont
  {Giovannetti}}, \bibinfo {author} {\bibfnamefont {A.~N.}\ \bibnamefont
  {Kato}},\ and\ \bibinfo {author} {\bibfnamefont {K.}~\bibnamefont {Yuasa}},\
  }\bibfield  {title} {\bibinfo {title} {Quantum estimation via sequential
  measurements},\ }\href {https://doi.org/10.1088/1367-2630/17/11/113055}
  {\bibfield  {journal} {\bibinfo  {journal} {New J. Phys.}\ }\textbf {\bibinfo
  {volume} {17}},\ \bibinfo {pages} {113055} (\bibinfo {year}
  {2015})}\BibitemShut {NoStop}%
\bibitem [{\citenamefont {De~Pasquale}\ \emph {et~al.}(2017)\citenamefont
  {De~Pasquale}, \citenamefont {Yuasa},\ and\ \citenamefont
  {Giovannetti}}]{DePasquale2017}%
  \BibitemOpen
  \bibfield  {author} {\bibinfo {author} {\bibfnamefont {A.}~\bibnamefont
  {De~Pasquale}}, \bibinfo {author} {\bibfnamefont {K.}~\bibnamefont {Yuasa}},\
  and\ \bibinfo {author} {\bibfnamefont {V.}~\bibnamefont {Giovannetti}},\
  }\bibfield  {title} {\bibinfo {title} {Estimating temperature via sequential
  measurements},\ }\href {https://doi.org/10.1103/PhysRevA.96.012316}
  {\bibfield  {journal} {\bibinfo  {journal} {Phys. Rev. A}\ }\textbf {\bibinfo
  {volume} {96}},\ \bibinfo {pages} {012316} (\bibinfo {year}
  {2017})}\BibitemShut {NoStop}%
\bibitem [{\citenamefont {Montenegro}\ \emph {et~al.}(2022)\citenamefont
  {Montenegro}, \citenamefont {Jones}, \citenamefont {Bose},\ and\
  \citenamefont {Bayat}}]{Montenegro2022}%
  \BibitemOpen
  \bibfield  {author} {\bibinfo {author} {\bibfnamefont {V.}~\bibnamefont
  {Montenegro}}, \bibinfo {author} {\bibfnamefont {G.~S.}\ \bibnamefont
  {Jones}}, \bibinfo {author} {\bibfnamefont {S.}~\bibnamefont {Bose}},\ and\
  \bibinfo {author} {\bibfnamefont {A.}~\bibnamefont {Bayat}},\ }\bibfield
  {title} {\bibinfo {title} {Sequential measurements for quantum-enhanced
  magnetometry in spin chain probes},\ }\href
  {https://doi.org/10.1103/PhysRevLett.129.120503} {\bibfield  {journal}
  {\bibinfo  {journal} {Phys. Rev. Lett.}\ }\textbf {\bibinfo {volume} {129}},\
  \bibinfo {pages} {120503} (\bibinfo {year} {2022})}\BibitemShut {NoStop}%
\bibitem [{\citenamefont {Radaelli}\ \emph {et~al.}(2023)\citenamefont
  {Radaelli}, \citenamefont {Landi}, \citenamefont {Modi},\ and\ \citenamefont
  {Binder}}]{Radaelli2023}%
  \BibitemOpen
  \bibfield  {author} {\bibinfo {author} {\bibfnamefont {M.}~\bibnamefont
  {Radaelli}}, \bibinfo {author} {\bibfnamefont {G.~T.}\ \bibnamefont {Landi}},
  \bibinfo {author} {\bibfnamefont {K.}~\bibnamefont {Modi}},\ and\ \bibinfo
  {author} {\bibfnamefont {F.~C.}\ \bibnamefont {Binder}},\ }\bibfield  {title}
  {\bibinfo {title} {Fisher information of correlated stochastic processes},\
  }\href {https://doi.org/10.1088/1367-2630/acd321} {\bibfield  {journal}
  {\bibinfo  {journal} {New J. Phys.}\ }\textbf {\bibinfo {volume} {25}},\
  \bibinfo {pages} {053037} (\bibinfo {year} {2023})}\BibitemShut {NoStop}%
\bibitem [{\citenamefont {Giovannetti}\ \emph {et~al.}(2006)\citenamefont
  {Giovannetti}, \citenamefont {Lloyd},\ and\ \citenamefont
  {Maccone}}]{Giovannetti2006}%
  \BibitemOpen
  \bibfield  {author} {\bibinfo {author} {\bibfnamefont {V.}~\bibnamefont
  {Giovannetti}}, \bibinfo {author} {\bibfnamefont {S.}~\bibnamefont {Lloyd}},\
  and\ \bibinfo {author} {\bibfnamefont {L.}~\bibnamefont {Maccone}},\
  }\bibfield  {title} {\bibinfo {title} {Quantum metrology},\ }\href
  {https://doi.org/10.1103/PhysRevLett.96.010401} {\bibfield  {journal}
  {\bibinfo  {journal} {Phys. Rev. Lett.}\ }\textbf {\bibinfo {volume} {96}},\
  \bibinfo {pages} {010401} (\bibinfo {year} {2006})}\BibitemShut {NoStop}%
\bibitem [{\citenamefont {Giovannetti}\ \emph {et~al.}(2011)\citenamefont
  {Giovannetti}, \citenamefont {Lloyd},\ and\ \citenamefont
  {Maccone}}]{Giovannetti2011}%
  \BibitemOpen
  \bibfield  {author} {\bibinfo {author} {\bibfnamefont {V.}~\bibnamefont
  {Giovannetti}}, \bibinfo {author} {\bibfnamefont {S.}~\bibnamefont {Lloyd}},\
  and\ \bibinfo {author} {\bibfnamefont {L.}~\bibnamefont {Maccone}},\
  }\bibfield  {title} {\bibinfo {title} {Advances in quantum metrology},\
  }\href {https://doi.org/10.1038/nphoton.2011.35} {\bibfield  {journal}
  {\bibinfo  {journal} {Nat. Photonics}\ }\textbf {\bibinfo {volume} {5}},\
  \bibinfo {pages} {222} (\bibinfo {year} {2011})}\BibitemShut {NoStop}%
\bibitem [{\citenamefont {Braun}\ \emph {et~al.}(2018)\citenamefont {Braun},
  \citenamefont {Adesso}, \citenamefont {Benatti}, \citenamefont {Floreanini},
  \citenamefont {Marzolino}, \citenamefont {Mitchell},\ and\ \citenamefont
  {Pirandola}}]{Braun2018}%
  \BibitemOpen
  \bibfield  {author} {\bibinfo {author} {\bibfnamefont {D.}~\bibnamefont
  {Braun}}, \bibinfo {author} {\bibfnamefont {G.}~\bibnamefont {Adesso}},
  \bibinfo {author} {\bibfnamefont {F.}~\bibnamefont {Benatti}}, \bibinfo
  {author} {\bibfnamefont {R.}~\bibnamefont {Floreanini}}, \bibinfo {author}
  {\bibfnamefont {U.}~\bibnamefont {Marzolino}}, \bibinfo {author}
  {\bibfnamefont {M.~W.}\ \bibnamefont {Mitchell}},\ and\ \bibinfo {author}
  {\bibfnamefont {S.}~\bibnamefont {Pirandola}},\ }\bibfield  {title} {\bibinfo
  {title} {Quantum-enhanced measurements without entanglement},\ }\href
  {https://doi.org/10.1103/RevModPhys.90.035006} {\bibfield  {journal}
  {\bibinfo  {journal} {Rev. Mod. Phys.}\ }\textbf {\bibinfo {volume} {90}},\
  \bibinfo {pages} {035006} (\bibinfo {year} {2018})}\BibitemShut {NoStop}%
\bibitem [{\citenamefont {Wang}\ \emph {et~al.}(2019)\citenamefont {Wang},
  \citenamefont {Chen}, \citenamefont {Peng}, \citenamefont {Wrachtrup},\ and\
  \citenamefont {Liu}}]{Wang2019}%
  \BibitemOpen
  \bibfield  {author} {\bibinfo {author} {\bibfnamefont {P.}~\bibnamefont
  {Wang}}, \bibinfo {author} {\bibfnamefont {C.}~\bibnamefont {Chen}}, \bibinfo
  {author} {\bibfnamefont {X.}~\bibnamefont {Peng}}, \bibinfo {author}
  {\bibfnamefont {J.}~\bibnamefont {Wrachtrup}},\ and\ \bibinfo {author}
  {\bibfnamefont {R.-B.}\ \bibnamefont {Liu}},\ }\bibfield  {title} {\bibinfo
  {title} {Characterization of arbitrary-order correlations in quantum baths by
  weak measurement},\ }\href {https://doi.org/10.1103/PhysRevLett.123.050603}
  {\bibfield  {journal} {\bibinfo  {journal} {Phys. Rev. Lett.}\ }\textbf
  {\bibinfo {volume} {123}},\ \bibinfo {pages} {050603} (\bibinfo {year}
  {2019})}\BibitemShut {NoStop}%
\bibitem [{\citenamefont {Wang}\ \emph {et~al.}(2021)\citenamefont {Wang},
  \citenamefont {Chen},\ and\ \citenamefont {Liu}}]{Wang2021}%
  \BibitemOpen
  \bibfield  {author} {\bibinfo {author} {\bibfnamefont {P.}~\bibnamefont
  {Wang}}, \bibinfo {author} {\bibfnamefont {C.}~\bibnamefont {Chen}},\ and\
  \bibinfo {author} {\bibfnamefont {R.-B.}\ \bibnamefont {Liu}},\ }\bibfield
  {title} {\bibinfo {title} {Classical-noise-free sensing based on quantum
  correlation measurement},\ }\href
  {https://doi.org/10.1088/0256-307X/38/1/010301} {\bibfield  {journal}
  {\bibinfo  {journal} {Chin. Phys. Lett.}\ }\textbf {\bibinfo {volume} {38}},\
  \bibinfo {pages} {010301} (\bibinfo {year} {2021})}\BibitemShut {NoStop}%
\bibitem [{\citenamefont {Wu}\ \emph {et~al.}(2024)\citenamefont {Wu},
  \citenamefont {Wang}, \citenamefont {Wang}, \citenamefont {Li}, \citenamefont
  {Liu}, \citenamefont {Chen}, \citenamefont {Peng},\ and\ \citenamefont
  {Liu}}]{Wu2024}%
  \BibitemOpen
  \bibfield  {author} {\bibinfo {author} {\bibfnamefont {Z.}~\bibnamefont
  {Wu}}, \bibinfo {author} {\bibfnamefont {P.}~\bibnamefont {Wang}}, \bibinfo
  {author} {\bibfnamefont {T.}~\bibnamefont {Wang}}, \bibinfo {author}
  {\bibfnamefont {Y.}~\bibnamefont {Li}}, \bibinfo {author} {\bibfnamefont
  {R.}~\bibnamefont {Liu}}, \bibinfo {author} {\bibfnamefont {Y.}~\bibnamefont
  {Chen}}, \bibinfo {author} {\bibfnamefont {X.}~\bibnamefont {Peng}},\ and\
  \bibinfo {author} {\bibfnamefont {R.-B.}\ \bibnamefont {Liu}},\ }\bibfield
  {title} {\bibinfo {title} {Selective detection of dynamics-complete set of
  correlations via quantum channels},\ }\href
  {https://doi.org/10.1103/PhysRevLett.132.200802} {\bibfield  {journal}
  {\bibinfo  {journal} {Phys. Rev. Lett.}\ }\textbf {\bibinfo {volume} {132}},\
  \bibinfo {pages} {200802} (\bibinfo {year} {2024})}\BibitemShut {NoStop}%
\bibitem [{\citenamefont {Cheung}\ and\ \citenamefont {Liu}()}]{Cheung2024}%
  \BibitemOpen
  \bibfield  {author} {\bibinfo {author} {\bibfnamefont {B.~C.~H.}\
  \bibnamefont {Cheung}}\ and\ \bibinfo {author} {\bibfnamefont {R.-B.}\
  \bibnamefont {Liu}},\ }\bibfield  {title} {\bibinfo {title} {Quantum
  nonlinear spectroscopy via correlations of weak faraday-rotation
  measurements},\ }\href
  {https://doi.org/https://doi.org/10.1002/qute.202300286} {\bibinfo  {journal}
  {Adv. Quantum Technol.}\ ,\ \bibinfo {pages} {2300286}}\BibitemShut {NoStop}%
\bibitem [{\citenamefont {Meinel}\ \emph {et~al.}(2022)\citenamefont {Meinel},
  \citenamefont {Vorobyov}, \citenamefont {Wang}, \citenamefont {Yavkin},
  \citenamefont {Pfender}, \citenamefont {Sumiya}, \citenamefont {Onoda},
  \citenamefont {Isoya}, \citenamefont {Liu},\ and\ \citenamefont
  {Wrachtrup}}]{Meinel2022}%
  \BibitemOpen
\bibfield  {journal} {  }\bibfield  {author} {\bibinfo {author} {\bibfnamefont
  {J.}~\bibnamefont {Meinel}}, \bibinfo {author} {\bibfnamefont
  {V.}~\bibnamefont {Vorobyov}}, \bibinfo {author} {\bibfnamefont
  {P.}~\bibnamefont {Wang}}, \bibinfo {author} {\bibfnamefont {B.}~\bibnamefont
  {Yavkin}}, \bibinfo {author} {\bibfnamefont {M.}~\bibnamefont {Pfender}},
  \bibinfo {author} {\bibfnamefont {H.}~\bibnamefont {Sumiya}}, \bibinfo
  {author} {\bibfnamefont {S.}~\bibnamefont {Onoda}}, \bibinfo {author}
  {\bibfnamefont {J.}~\bibnamefont {Isoya}}, \bibinfo {author} {\bibfnamefont
  {R.-B.}\ \bibnamefont {Liu}},\ and\ \bibinfo {author} {\bibfnamefont
  {J.}~\bibnamefont {Wrachtrup}},\ }\bibfield  {title} {\bibinfo {title}
  {Quantum nonlinear spectroscopy of single nuclear spins},\ }\href
  {https://doi.org/10.1038/s41467-022-32610-8} {\bibfield  {journal} {\bibinfo
  {journal} {Nat. Commun.}\ }\textbf {\bibinfo {volume} {13}},\ \bibinfo
  {pages} {5318} (\bibinfo {year} {2022})}\BibitemShut {NoStop}%
\bibitem [{\citenamefont {Shen}\ \emph {et~al.}(2023)\citenamefont {Shen},
  \citenamefont {Wang}, \citenamefont {Cheung}, \citenamefont {Wrachtrup},
  \citenamefont {Liu},\ and\ \citenamefont {Yang}}]{Shen2023}%
  \BibitemOpen
  \bibfield  {author} {\bibinfo {author} {\bibfnamefont {Y.}~\bibnamefont
  {Shen}}, \bibinfo {author} {\bibfnamefont {P.}~\bibnamefont {Wang}}, \bibinfo
  {author} {\bibfnamefont {C.~T.}\ \bibnamefont {Cheung}}, \bibinfo {author}
  {\bibfnamefont {J.}~\bibnamefont {Wrachtrup}}, \bibinfo {author}
  {\bibfnamefont {R.-B.}\ \bibnamefont {Liu}},\ and\ \bibinfo {author}
  {\bibfnamefont {S.}~\bibnamefont {Yang}},\ }\bibfield  {title} {\bibinfo
  {title} {Detection of quantum signals free of classical noise via quantum
  correlation},\ }\href {https://doi.org/10.1103/PhysRevLett.130.070802}
  {\bibfield  {journal} {\bibinfo  {journal} {Phys. Rev. Lett.}\ }\textbf
  {\bibinfo {volume} {130}},\ \bibinfo {pages} {070802} (\bibinfo {year}
  {2023})}\BibitemShut {NoStop}%
\bibitem [{\citenamefont {Laraoui}\ \emph {et~al.}(2013)\citenamefont
  {Laraoui}, \citenamefont {Dolde}, \citenamefont {Burk}, \citenamefont
  {Reinhard}, \citenamefont {Wrachtrup},\ and\ \citenamefont
  {Meriles}}]{Laraoui2013}%
  \BibitemOpen
  \bibfield  {author} {\bibinfo {author} {\bibfnamefont {A.}~\bibnamefont
  {Laraoui}}, \bibinfo {author} {\bibfnamefont {F.}~\bibnamefont {Dolde}},
  \bibinfo {author} {\bibfnamefont {C.}~\bibnamefont {Burk}}, \bibinfo {author}
  {\bibfnamefont {F.}~\bibnamefont {Reinhard}}, \bibinfo {author}
  {\bibfnamefont {J.}~\bibnamefont {Wrachtrup}},\ and\ \bibinfo {author}
  {\bibfnamefont {C.~A.}\ \bibnamefont {Meriles}},\ }\bibfield  {title}
  {\bibinfo {title} {High-resolution correlation spectroscopy of 13c spins near
  a nitrogen-vacancy centre in diamond},\ }\href
  {https://doi.org/10.1038/ncomms2685} {\bibfield  {journal} {\bibinfo
  {journal} {Nat. Commun.}\ }\textbf {\bibinfo {volume} {4}},\ \bibinfo {pages}
  {1651} (\bibinfo {year} {2013})}\BibitemShut {NoStop}%
\bibitem [{\citenamefont {Aslam}\ \emph {et~al.}(2017)\citenamefont {Aslam},
  \citenamefont {Pfender}, \citenamefont {Neumann}, \citenamefont {Reuter},
  \citenamefont {Zappe}, \citenamefont {de~Oliveira}, \citenamefont
  {Denisenko}, \citenamefont {Sumiya}, \citenamefont {Onoda}, \citenamefont
  {Isoya},\ and\ \citenamefont {Wrachtrup}}]{Aslam2017}%
  \BibitemOpen
  \bibfield  {author} {\bibinfo {author} {\bibfnamefont {N.}~\bibnamefont
  {Aslam}}, \bibinfo {author} {\bibfnamefont {M.}~\bibnamefont {Pfender}},
  \bibinfo {author} {\bibfnamefont {P.}~\bibnamefont {Neumann}}, \bibinfo
  {author} {\bibfnamefont {R.}~\bibnamefont {Reuter}}, \bibinfo {author}
  {\bibfnamefont {A.}~\bibnamefont {Zappe}}, \bibinfo {author} {\bibfnamefont
  {F.~F.}\ \bibnamefont {de~Oliveira}}, \bibinfo {author} {\bibfnamefont
  {A.}~\bibnamefont {Denisenko}}, \bibinfo {author} {\bibfnamefont
  {H.}~\bibnamefont {Sumiya}}, \bibinfo {author} {\bibfnamefont
  {S.}~\bibnamefont {Onoda}}, \bibinfo {author} {\bibfnamefont
  {J.}~\bibnamefont {Isoya}},\ and\ \bibinfo {author} {\bibfnamefont
  {J.}~\bibnamefont {Wrachtrup}},\ }\bibfield  {title} {\bibinfo {title}
  {Nanoscale nuclear magnetic resonance with chemical resolution},\ }\href
  {https://doi.org/10.1126/science.aam8697} {\bibfield  {journal} {\bibinfo
  {journal} {Science}\ }\textbf {\bibinfo {volume} {357}},\ \bibinfo {pages}
  {67} (\bibinfo {year} {2017})}\BibitemShut {NoStop}%
\bibitem [{\citenamefont {Boss}\ \emph {et~al.}(2017)\citenamefont {Boss},
  \citenamefont {Cujia}, \citenamefont {Zopes},\ and\ \citenamefont
  {Degen}}]{Boss2017}%
  \BibitemOpen
  \bibfield  {author} {\bibinfo {author} {\bibfnamefont {J.~M.}\ \bibnamefont
  {Boss}}, \bibinfo {author} {\bibfnamefont {K.~S.}\ \bibnamefont {Cujia}},
  \bibinfo {author} {\bibfnamefont {J.}~\bibnamefont {Zopes}},\ and\ \bibinfo
  {author} {\bibfnamefont {C.~L.}\ \bibnamefont {Degen}},\ }\bibfield  {title}
  {\bibinfo {title} {Quantum sensing with arbitrary frequency resolution},\
  }\href {https://doi.org/10.1126/science.aam7009} {\bibfield  {journal}
  {\bibinfo  {journal} {Science}\ }\textbf {\bibinfo {volume} {356}},\ \bibinfo
  {pages} {837} (\bibinfo {year} {2017})}\BibitemShut {NoStop}%
\bibitem [{\citenamefont {Glenn}\ \emph {et~al.}(2018)\citenamefont {Glenn},
  \citenamefont {Bucher}, \citenamefont {Lee}, \citenamefont {Lukin},
  \citenamefont {Park},\ and\ \citenamefont {Walsworth}}]{Glenn2018}%
  \BibitemOpen
  \bibfield  {author} {\bibinfo {author} {\bibfnamefont {D.~R.}\ \bibnamefont
  {Glenn}}, \bibinfo {author} {\bibfnamefont {D.~B.}\ \bibnamefont {Bucher}},
  \bibinfo {author} {\bibfnamefont {J.}~\bibnamefont {Lee}}, \bibinfo {author}
  {\bibfnamefont {M.~D.}\ \bibnamefont {Lukin}}, \bibinfo {author}
  {\bibfnamefont {H.}~\bibnamefont {Park}},\ and\ \bibinfo {author}
  {\bibfnamefont {R.~L.}\ \bibnamefont {Walsworth}},\ }\bibfield  {title}
  {\bibinfo {title} {High-resolution magnetic resonance spectroscopy using a
  solid-state spin sensor},\ }\href {https://doi.org/10.1038/nature25781}
  {\bibfield  {journal} {\bibinfo  {journal} {Nature}\ }\textbf {\bibinfo
  {volume} {555}},\ \bibinfo {pages} {351} (\bibinfo {year}
  {2018})}\BibitemShut {NoStop}%
\bibitem [{\citenamefont {Pfender}\ \emph {et~al.}(2019)\citenamefont
  {Pfender}, \citenamefont {Wang}, \citenamefont {Sumiya}, \citenamefont
  {Onoda}, \citenamefont {Yang}, \citenamefont {Dasari}, \citenamefont
  {Neumann}, \citenamefont {Pan}, \citenamefont {Isoya}, \citenamefont {Liu},\
  and\ \citenamefont {Wrachtrup}}]{Pfender2019}%
  \BibitemOpen
  \bibfield  {author} {\bibinfo {author} {\bibfnamefont {M.}~\bibnamefont
  {Pfender}}, \bibinfo {author} {\bibfnamefont {P.}~\bibnamefont {Wang}},
  \bibinfo {author} {\bibfnamefont {H.}~\bibnamefont {Sumiya}}, \bibinfo
  {author} {\bibfnamefont {S.}~\bibnamefont {Onoda}}, \bibinfo {author}
  {\bibfnamefont {W.}~\bibnamefont {Yang}}, \bibinfo {author} {\bibfnamefont
  {D.~B.~R.}\ \bibnamefont {Dasari}}, \bibinfo {author} {\bibfnamefont
  {P.}~\bibnamefont {Neumann}}, \bibinfo {author} {\bibfnamefont {X.-Y.}\
  \bibnamefont {Pan}}, \bibinfo {author} {\bibfnamefont {J.}~\bibnamefont
  {Isoya}}, \bibinfo {author} {\bibfnamefont {R.-B.}\ \bibnamefont {Liu}},\
  and\ \bibinfo {author} {\bibfnamefont {J.}~\bibnamefont {Wrachtrup}},\
  }\bibfield  {title} {\bibinfo {title} {High-resolution spectroscopy of single
  nuclear spins via sequential weak measurements},\ }\href
  {https://doi.org/10.1038/s41467-019-08544-z} {\bibfield  {journal} {\bibinfo
  {journal} {Nat. Commun.}\ }\textbf {\bibinfo {volume} {10}},\ \bibinfo
  {pages} {594} (\bibinfo {year} {2019})}\BibitemShut {NoStop}%
\bibitem [{\citenamefont {Cujia}\ \emph {et~al.}(2019)\citenamefont {Cujia},
  \citenamefont {Boss}, \citenamefont {Herb}, \citenamefont {Zopes},\ and\
  \citenamefont {Degen}}]{Cujia2019}%
  \BibitemOpen
  \bibfield  {author} {\bibinfo {author} {\bibfnamefont {K.~S.}\ \bibnamefont
  {Cujia}}, \bibinfo {author} {\bibfnamefont {J.~M.}\ \bibnamefont {Boss}},
  \bibinfo {author} {\bibfnamefont {K.}~\bibnamefont {Herb}}, \bibinfo {author}
  {\bibfnamefont {J.}~\bibnamefont {Zopes}},\ and\ \bibinfo {author}
  {\bibfnamefont {C.~L.}\ \bibnamefont {Degen}},\ }\bibfield  {title} {\bibinfo
  {title} {Tracking the precession of single nuclear spins by weak
  measurements},\ }\href {https://doi.org/10.1038/s41586-019-1334-9} {\bibfield
   {journal} {\bibinfo  {journal} {Nature}\ }\textbf {\bibinfo {volume}
  {571}},\ \bibinfo {pages} {230} (\bibinfo {year} {2019})}\BibitemShut
  {NoStop}%
\bibitem [{\citenamefont {Kovachy}\ \emph {et~al.}(2015)\citenamefont
  {Kovachy}, \citenamefont {Hogan}, \citenamefont {Sugarbaker}, \citenamefont
  {Dickerson}, \citenamefont {Donnelly}, \citenamefont {Overstreet},\ and\
  \citenamefont {Kasevich}}]{Kovachy2015}%
  \BibitemOpen
  \bibfield  {author} {\bibinfo {author} {\bibfnamefont {T.}~\bibnamefont
  {Kovachy}}, \bibinfo {author} {\bibfnamefont {J.~M.}\ \bibnamefont {Hogan}},
  \bibinfo {author} {\bibfnamefont {A.}~\bibnamefont {Sugarbaker}}, \bibinfo
  {author} {\bibfnamefont {S.~M.}\ \bibnamefont {Dickerson}}, \bibinfo {author}
  {\bibfnamefont {C.~A.}\ \bibnamefont {Donnelly}}, \bibinfo {author}
  {\bibfnamefont {C.}~\bibnamefont {Overstreet}},\ and\ \bibinfo {author}
  {\bibfnamefont {M.~A.}\ \bibnamefont {Kasevich}},\ }\bibfield  {title}
  {\bibinfo {title} {Matter wave lensing to picokelvin temperatures},\ }\href
  {https://doi.org/10.1103/PhysRevLett.114.143004} {\bibfield  {journal}
  {\bibinfo  {journal} {Phys. Rev. Lett.}\ }\textbf {\bibinfo {volume} {114}},\
  \bibinfo {pages} {143004} (\bibinfo {year} {2015})}\BibitemShut {NoStop}%
\bibitem [{\citenamefont {Breuer}\ and\ \citenamefont
  {Petruccione}(2007)}]{Breuer2007}%
  \BibitemOpen
  \bibfield  {author} {\bibinfo {author} {\bibfnamefont {H.-P.}\ \bibnamefont
  {Breuer}}\ and\ \bibinfo {author} {\bibfnamefont {F.}~\bibnamefont
  {Petruccione}},\ }\href@noop {} {\emph {\bibinfo {title} {The Theory of Open
  Quantum Systems}}}\ (\bibinfo  {publisher} {{Oxford University Press}},\
  \bibinfo {year} {2007})\BibitemShut {NoStop}%
\bibitem [{\citenamefont {Degen}\ \emph {et~al.}(2017)\citenamefont {Degen},
  \citenamefont {Reinhard},\ and\ \citenamefont {Cappellaro}}]{Degen2017}%
  \BibitemOpen
  \bibfield  {author} {\bibinfo {author} {\bibfnamefont {C.~L.}\ \bibnamefont
  {Degen}}, \bibinfo {author} {\bibfnamefont {F.}~\bibnamefont {Reinhard}},\
  and\ \bibinfo {author} {\bibfnamefont {P.}~\bibnamefont {Cappellaro}},\
  }\bibfield  {title} {\bibinfo {title} {Quantum sensing},\ }\href
  {https://doi.org/10.1103/RevModPhys.89.035002} {\bibfield  {journal}
  {\bibinfo  {journal} {Rev. Mod. Phys.}\ }\textbf {\bibinfo {volume} {89}},\
  \bibinfo {pages} {035002} (\bibinfo {year} {2017})}\BibitemShut {NoStop}%
\end{thebibliography}%

\end{document}